\documentclass[12pt]{amsart}

\usepackage{color} 
\usepackage{graphicx} 
\usepackage{latexsym,amssymb,amsmath,amsfonts} 
\usepackage{mathrsfs} 
\usepackage{fixltx2e} 
\usepackage{tikz} 
\usetikzlibrary{shapes,patterns,arrows,decorations.pathreplacing}
\usepackage{epigraph}

\allowdisplaybreaks[4]
\usepackage{latexsym, amssymb, amsmath}
\usepackage{enumerate}
\usepackage{flushend} 
\usepackage{mathrsfs} 
\usepackage{stfloats}
\def\custombibliography#1{
 \normalsize
\section*{\centering References}
 \list
 {[\arabic{enumi}]}{\settowidth\labelwidth{[#1]}\leftmargin\labelwidth
 \setlength{\itemsep}{.1em}
 \advance\leftmargin\labelsep
 \usecounter{enumi}}
 \def\newblock{\hskip .11em plus .33em minus -.07em}
 \sloppy
 \sfcode`\.=1000\relax}

\def\L2{{\cal L}_2}

\newcommand\bull{\vrule height .9ex width .8ex depth -.1ex } 

\newcommand\re{\rm I\! R}
\newcommand\cdcout[1]{} 

\newcommand{\rv}[1]{\boldsymbol{#1}} 
\newcommand{\RomanNumber}[1]{\uppercase\expandafter{\romannumeral #1}}
\newcommand{\romannumber}[1]{\lowercase\expandafter{\romannumeral #1}}

\DeclareMathAlphabet{\mathpzc}{OT1}{pzc}{m}{it}

\def\1{\rv 1} 

\usepackage{flushend} 
\usepackage{stfloats}
\usepackage{color} 
\usepackage{latexsym,amssymb,amsmath} 
\usepackage{mathrsfs,mathtools} 
\usepackage{multicol,enumerate}

\def\abs#1{\lvert #1 \rvert}

\def\allcommutingpolyX#1{\mbox{$\re^{#1}\,[\tilde{X}]$}}

\def\allcommutingseries{\mbox{$\re\, [[\tilde{X}]]$}}

\def\allcommutingseriessingvar{\mbox{$\re\,[[\tilde{x}_{1}]]$}}
\def\allcommutingpolysingvar{\mbox{$\re\,[\tilde{x}_{1}]$}}

\def\allcommutingseriesXell{\mbox{$\re^{\ell}\, [[X]]$}} 

\def\allcommutingseriesX#1{\mbox{$\re^{#1}\, [[\tilde{X}]]$}}
\def\allcommutingseriesLCX#1{\mbox{$\re_{LC}^{#1}\, [[\tilde{X}]]$}}
\def\allcommutingseriesGCX#1{\mbox{$\re_{GC}^{#1}\, [[\tilde{X}]]$}}

\def\allpoly{\mbox{$\re\langle X \rangle$}}
\def\allpolyell{\mbox{$\re^{\ell}\langle X \rangle$}}
\def\allpolyx0degn{\mbox{$P_n$}}

\def\allwords{\mbox{$X^{\ast}$}}
\def\allcommutingwords{\mbox{$\tilde{X}^{\ast}$}}
\def\allseries{\mbox{$\re\langle\langle X \rangle\rangle$}}
\def\allseries#1{\mbox{$\re^{#1}\langle\langle X \rangle\rangle$}}
\def\allseriesXpri#1{\mbox{$\re^{#1}\langle\langle X^{\prime} \rangle\rangle$}}
\def\allseriesdelta{\mbox{$\delta + \re\langle\langle X \rangle\rangle$}}
\def\allseriesdelta#1{\mbox{$\delta + \re^{#1}\langle\langle X \rangle\rangle$}}
\def\allseriesell{\mbox{$\re^{\ell} \langle\langle X \rangle\rangle$}}
\def\allseriesGC#1{\mbox{$\re^{#1}_{GC}\langle\langle X \rangle\rangle$}}

\def\allseriesLC#1{\mbox{$\re^{#1}_{LC}\langle\langle X \rangle\rangle$}}
\def\allseriesLCXtil#1{\mbox{$\re^{#1}_{LC}\langle\langle \tilde{X} \rangle\rangle$}}
\def\allseriespLC#1{\mbox{$\re^{#1}_{p,LC}\langle\langle X \rangle\rangle$}}

\def\allseriesmLC{\mbox{$\re^{m}_{LC}\langle\langle X \rangle\rangle$}}

\def\allproperseriesell{\mbox{$\re_{p}^{\ell}\, \langle\langle X \rangle\rangle$}}
\def\allproperseries#1{\mbox{$\re_{p}^{#1}\, \langle\langle X \rangle\rangle$}}

\def\allseriesellLC{\mbox{$\re^{\ell}_{LC}\langle\langle X \rangle\rangle$}}
\def\allseriesellpriLC{\mbox{$\re^{\ell}_{LC}\langle\langle X^{\prime} \rangle\rangle$}}

\def\allseriesX1{\mbox{$\re [[ X_1 ]]$}}
\def\serSinf#1{\mbox{$S_{\infty}^{#1}$}}
\def\serSinfR#1#2{\mbox{$S_{\infty}^{#1}(#2)$}}
\def\bull{\rule{0.08in}{0.08in}} 

\newcommand{\comment}[1]{} 

\def\doubleone{{\rm\, l\!l}}

\def\Endallseries{{\rm End}\left(\allseries{}\right)}

\def\eqref#1{(\ref{#1})} 

\def\Fliessdelta{\mathscr{F}_{\delta}}

\def\mbf#1{\hbox{\mathversion{bold}$#1$}} 
\def\modcomp{\:\tilde{\circ}\,} 
\def\mixcomp{\:\hat{\circ}\,} 
\def\nat{{\mathbb N}} 
\def\norm#1{\Vert#1\Vert}
\def\notin{{\not\in}}

\def\norminf#1#2{{\left\lvert\left\lvert#2\right\rvert\right\rvert}_{\infty,#1}}

\def\openbull{\framebox[0.08in][c]{$\;$}} 

\def\re{{\mathbb R}} 

\def\shuffle{{\scriptscriptstyle \;\sqcup \hspace*{-0.05cm}\sqcup\;}}

\def\supp{{\rm supp}}
\def\orde{{\rm ord}}

\def\begals{\[\begin{aligned}}
\def\endals{\end{aligned}\]}
\def\begal{\begin{align*}}
\def\endal{\end{align*}}
\def\begce{\begin{center}}
\def\endce{\end{center}}
\def\begar{\begin{array}}
\def\endar{\end{array}}
\def\begeq{\begin{equation}}
\def\endeq{\end{equation}}
\def\begdi{\begin{displaymath}}
\def\enddi{\end{displaymath}}
\def\begdis{\begin{eqnarray*}}
\def\enddis{\end{eqnarray*}}
\def\begeqa{\begin{eqnarray}}
\def\endeqa{\end{eqnarray}}
\def\begdes{\begin{description}}
\def\enddes{\end{description}}
\def\begit{\begin{itemize}}
\def\endit{\end{itemize}}
\def\begen{\begin{enumerate}}
\def\enden{\end{enumerate}}
\def\beglar{\left[\begin{array}}
\def\endrar{\end{array}\right]}
\def\begle{\begin{mylemma}}
\def\endle{\end{mylemma}}
\def\begde{\begin{mydefinition}}
\def\endde{\end{mydefinition}}
\def\begth{\begin{mytheorem}}
\def\endth{\end{mytheorem}}
\def\begco{\begin{mycorollary}}
\def\endco{\end{mycorollary}}
\def\begprop{\begin{myproposition}}
\def\endprop{\end{myproposition}}
\def\begex{\begin{myexample}}
\def\endex{\hfill\openbull \end{myexample}}
\def\begexer{\begin{myexercise}}
\def\endexer{\end{myexercise}}
\def\begalg{\begin{myalgo}}
\def\endalg{\end{myalgo}}

\def\begres{\noindent{\bf Remarks}:\begin{enumerate}}
\def\endres{\end{enumerate} \par}
\def\begpr{\noindent{\em Proof:}$\;\;$}
\def\endpr{\hfill\bull}
\def\begtab{\begin{tabular}}
\def\endtab{\end{tabular}}
\def\rref#1{(\ref{#1})}
\newtheorem{mylemma}{Lemma}[section]
\newtheorem{mydefinition}{Definition}[section]
\newtheorem{mytheorem}{Theorem}[section]
\newtheorem{mycorollary}{Corollary}[section]
\newtheorem{myexample}{Example}[section]
\newtheorem{myalgo}{Algorithm}[section]
\def\HAprod{\mbf{m}} 
\def\shuff#1#2{\mathbin{
      \hbox{\vbox{\hbox{\vrule \hskip#2 \vrule height#1 width 0pt}\hrule}\vbox{\hbox{\vrule \hskip#2 \vrule height#1 width 0pt\vrule }\hrule}}}}
\def\shuffl{{\mathchoice{\shuff{5pt}{3.5pt}}{\shuff{5pt}{3.5pt}}{\shuff{3pt}{2.6pt}}{\shuff{3pt}{2.6pt}}}}
\def\shuffle{{\, \shuffl \,}}

\begin{document}

\setlength{\emergencystretch}{2.3em} 

\title[Nonlinear Systems with Additive Static Feedback]{Formal Power Series Approach to Nonlinear Systems with Additive Static Feedback}
\author{G.~S.~Venkatesh\textsuperscript{$\dagger$}}
\thanks{$\dagger$Corresponding Author}
\address{Department of Electrical and Computer Engineering, Old Dominion University, Norfolk, Virginia 23529, USA}
\email{sgugg001@odu.edu}
\author{W.~Steven Gray}
\address{Department of Electrical and Computer Engineering, Old Dominion University, Norfolk, Virginia 23529, USA}
\email{sgray@odu.edu}
\urladdr{}
\date{\today}

\maketitle 

\begin{abstract}
	The goal of this paper is to compute the generating series of a closed-loop system when the plant is described in terms of a Chen-Fliess series and an additive static output feedback is applied. The first step is to consider the so called Wiener-Fliess connection consisting of a Chen-Fliess series followed by a memoryless function. Of particular importance will be the contractive nature of this map, which is needed to show that the closed-loop system has a Chen-Fliess series representation. To explicitly compute the generating series, two Hopf algebras are needed, the existing output feedback Hopf algebra used to describe dynamic output feedback, and the Hopf algebra of the shuffle group. These two combinatorial structures are combined to compute what will be called the Wiener-Fliess feedback product. It will be shown that this product has a natural interpretation as a transformation group acting on the plant and preserves the relative degree of the plant. The convergence of the Wiener-Fliess composition product and the additive static feedback product are completely characterized. 
\end{abstract}

\section{Introduction}
Let $F_c$ and $F_d$ be two nonlinear input-output systems represented by Chen-Fliess functional series \cite{Fliess_81}. It was shown in \cite{Ferfera_79,Gray-Li_05} that the feedback interconnection of two such systems always renders a closed-loop system in the same class. Its corresponding generating series, written as the {\em feedback product} $c@d$, can be efficiently computed in terms of a combinatorial Hopf algebra which is commutative, graded and connected \cite{Duffaut-Espinosa-etal_JA16,Foissy_15,Gray-etal_SCL14}. Convergence of the closed-loop system was characterized in detail by \cite{Thitsa-Gray_SIAM12}. Variations of the feedback product were used to solve system inversion problems \cite{Gray-etal_AUTO14} and trajectory generation problems \cite{Duffaut-Espinosa-Gray_ICSTCC17}.

What does not fit so neatly into this existing framework is the important case where the dynamical system $F_d$ in the feedback path is replaced with a memoryless function $f_d$, the so called static output feedback connection. The central problem here is that the loop contains a cascade connection of a Chen-Fliess series and a memoryless function, an object with an algebraic nature not entirely compatible with the algebras used to analyze the dynamic feedback case. Therefore, the goal of this paper is to address this problem by showing how to adapt existing algebraic tools for the analysis of static feedback systems. The first step is to reconsider the so called Wiener-Fliess connection consisting of a Chen-Fliess series followed by a memoryless function \cite{Gray_Thitsa_IJC12}. Of particular importance will be the contractive nature of this map, which is needed to show that the closed-loop system has a Chen-Fliess series representation. The preservation of relative degree under Wiener-Fliess composition is also described. Next, the focus turns to computing the generating series of the closed-loop system. What is needed in this regard are {\em two} Hopf algebras, the output feedback Hopf algebra described above, and the Hopf algebra of the shuffle group. These two combinatorial structures will be combined to compute what will be called the {\em Wiener-Fliess feedback product}. It will be shown that this product has a natural interpretation as a transformation group acting on the plant and preserves the relative degree of the plant. It should be noted that part of this work has already appeared in~\cite{GS-Gray_MTNS21}, where the focus was on static feedback restricted to the case of proper series. The restriction is lifted in the current work. In addition, the local and global convergence of the Wiener-Fliess composition product is characterized in detail. Finally, the local convergence of the Wiener-Fliess feedback product is proved, and a simple counterexample is provided to show that static feedback does not preserve global convergence. These results were developed as part of the doctoral dissertation of the first author \cite{GS_thesis}.

The paper is organized as follows. The next section provides a summary of the concepts related to Chen-Fliess series and their interconnections. The section also presents the necessary topological preliminaries. Section~\ref{sec:Wiener-Fliess} characterizes the Wiener-Fliess cascade interconnection. Section~\ref{sec:HA-shuffle-group} describes the Hopf algebra of the shuffle group and details regarding the computational framework. The static feedback connection is analyzed in Section~\ref{sec:static-feedback}.
The local convergence of the Wiener-Fliess composition product is extensively addressed in Section~\ref{sec:loc_conv}. The global convergence of the Wiener-Fliess composition product is established in Section~\ref{sec:glo_conv}. The nature of convergence of the additive static feedback is characterized in Section~\ref{sec:static_feedback}. The conclusions of the paper are given in the last section.

\section{Preliminaries}

A finite nonempty set of noncommuting symbols $X=\{ x_0,x_1,\ldots,x_m\}$ is called an {\em alphabet}. Each element of $X$ is called a {\em letter}, and any finite sequence of letters from $X$, $\eta=x_{i_1}\cdots x_{i_k}$, is called a {\em word} over $X$. Its {\em length} is $\abs{\eta}=k$. In particular, $\abs{\eta}_{x_i}$ is the number of times the letter $x_i\in X$ appears in $\eta$. The set of all words including the empty word, $\emptyset$, is denoted by $X^\ast$, and $X^+:=X^\ast\backslash\emptyset$. The set $X^\ast$ forms a monoid under catenation. The set of all words with prefix $\eta$ is written as $\eta X^\ast$. Any mapping $c:X^\ast\rightarrow\re^\ell$ is called a {\em formal power series}. The value of $c$ at $\eta\in X^\ast$ is denoted by $(c,\eta)$ and called the {\em coefficient} of $\eta$ in $c$. A series $c$ is {\em proper} when $(c,\emptyset)=0$. The {\em support} of $c$, $\supp(c)$, is the set of all words having nonzero coefficients. The {\em order} of $c$, $\orde(c)$, is the length of the minimal length word in its support. Normally, $c$ is written as a formal sum $c=\sum_{\eta\in X^\ast}(c,\eta)\eta.$ The collection of all formal power series over $X$ is denoted by $\allseriesell$. The set $\allseries{}$ is equipped with the partial ordering $\leq$ defined as : $c \leq d$ if and only $\abs{\left(c,\eta\right)} \leq \abs{\left(d,\eta\right)} \; \forall \eta \in \allwords$. A polynomial is a formal power series with finite support. The set of all noncommutative polynomials with coefficients in $\re^{\ell}$ is denoted by $\allpolyell$. The set of formal series $\allseriesell$ constitutes an associative $\re$-algebra under the catenation product and an associative and commutative $\re$-algebra under the {\em shuffle product}, that is, the bilinear product uniquely specified
by the shuffle product of two words
\begin{align}\label{eqn:shuff_prod_def}
(x_i\eta)\shuffle(x_j\xi)=x_i(\eta\shuffle(x_j\xi))+x_j((x_i\eta)\shuffle \xi),
\end{align}
where $x_i,x_j\in X$, $\eta,\xi\in X^\ast$ and with $\eta\shuffle\emptyset=\emptyset\shuffle\eta=\eta$ \cite{Fliess_81}. The subset of proper formal power series in $\allseriesell$ is denoted by $\allproperseries{\ell}$.

\begde\cite{Gray-etal_AUTO14,Gray_GS_relative degree} Let $X = \{x_0,x_1\}$. A generating series $c \in \allseries{}$ has relative degree $r$ if and only if there exists some $e\in\allseries{}$ with $x_1\not\in\supp(e)$ such that $c = c_N + Kx_0^{r-1}x_1 + x_0^{r-1}e$, where $c_N:=\sum_{k\geq 0} (c,x_0^k)x_0^k$ is called the \emph{natural part} of the series and $K\neq 0$. The relative degree of the series $c$ is denoted as $r_c$.
\endde
The set $\allseriesell$ is an ultrametric space with the ultrametric
\begals
\kappa(c,d) = \sigma^{\orde(c-d)},
\endals
where $c,d \in \allseriesell$ and $\sigma \in \: ]0,1[$. For brevity, $\kappa(c,0)$ is written as $\kappa(c)$, and $\kappa(c,d) = \kappa(c-d)$. The ultrametric space $(\allseriesell,\kappa)$ is Cauchy complete~\cite{Berstel-Reutenauer_88}. The following notions of \textit{strong} and \textit{weak} contraction maps will be used.

\begde Given metric spaces $(E,d)$ and $(E^{\prime},d^{\prime})$, a map $f : E \longrightarrow E^{\prime}$  is said to be a strong contraction map if $\forall s,t \in E$ it satisfies the condition $d^{\prime}(f(s),f(t)) \leq \alpha d(s,t)$, where $\alpha \in [0,1[$. If $\alpha = 1$, then the map $f$ is said to be a weak contraction map or a non-expansive map.
\endde

In the event that the letters of $X$ commute, the set of all corresponding formal power series is
denoted by $\allcommutingseriesXell$. For any series $c \in \allcommutingseriesXell$, the natural
number $\overline{\omega}(c)$ corresponds to the order of its proper part, namely, $c-(c,\emptyset)$.

\subsection{Fliess Operators}
\paragraph*{}
Let $\mathfrak{p}\ge 1$ and $t_0 < t_1$ be given. For a Lebesgue measurable
function $u: [t_0,t_1] \rightarrow\re^m$, define
$\norm{u}_{\mathfrak{p}}=\max\{\norm{u_i}_{\mathfrak{p}}: \ 1\le
i\le m\}$, where $\norm{u_i}_{\mathfrak{p}}$ is the usual
$L_{\mathfrak{p}}$-norm for a measurable real-valued function,
$u_i$, defined on $[t_0,t_1]$.  Let $L^m_{\mathfrak{p}}[t_0,t_1]$
denote the set of all measurable functions defined on $[t_0,t_1]$
having a finite $\norm{\cdot}_{\mathfrak{p}}$ norm and
$B_{\mathfrak{p}}^m(R)[t_0,t_1]:=\{u\in
L_{\mathfrak{p}}^m[t_0,t_1]:\norm{u}_{\mathfrak{p}}\leq R\}$. Assume $C[t_0,t_1]$
is the subset of continuous functions in $L_{1}^m[t_0,t_1]$.
Given any series $c\in\allseriesell$, the corresponding
{\em Chen-Fliess series} is
\begeq
F_c[u](t) =
\sum_{\eta\in X^{\ast}} (c,\eta)\,E_\eta[u](t,t_0), \label{eq:Fliess-operator-defined}
\endeq
where $E_\emptyset[u]=1$ and
\begdi
E_{x_i\bar{\eta}}[u](t,t_0) =
\int_{t_0}^tu_{i}(\tau)E_{\bar{\eta}}[u](\tau,t_0)\,d\tau
\enddi
with $x_i\in X$, $\bar{\eta}\in X^{\ast}$, and $u_0=1$ \cite{Fliess_81}.
If there exist constants $K,M>0$ such that
\begin{align}
\abs{(c_i,\eta)}\leq K M^{\abs{\eta}} \abs{\eta}!,\;\; \forall \eta\in X^\ast, \; \forall i = 1,\ldots,\ell\;,\label{eq:Gevrey-growth-condition}
\end{align}
then $F_c$ constitutes a well-defined mapping from
$B_{\mathfrak p}^m(R)[t_0,$ $t_0+T]$ into $B_{\mathfrak q}^{\ell}(S)[t_0, \, t_0+T]$ for sufficiently small $R,T >0$,
where the numbers $\mathfrak{p},\mathfrak{q}\in[1,\infty]$ are conjugate exponents, i.e., $1/\mathfrak{p}+1/\mathfrak{q}=1$~\cite{Gray-Wang_SCL02}. This map is referred to as a {\em Fliess operator}. A series $c \in \allseriesell$ obeying the growth condition in ~\rref{eq:Gevrey-growth-condition} is called a {\em locally convergent} generating series. The set of all locally convergent generating series is denoted by $\allseriesellLC$. The supremum of the set of all $\max\{R,T\}$ for which a Fliess operator $F_c$ is a well-defined mapping from $B_{\mathfrak p}^m(R)[t_0,$ $t_0+T]$ into $B_{\mathfrak a}^{\ell}(S)[t_0, \, t_0+T]$ is called the {\em radius of convergence} of the Fliess operator $F_c$ and is denoted by $\rho\left(F_c\right)$. A Fliess operator $F_c$ is called {\em locally convergent} if $\rho\left(F_c\right) > 0$. If there exist constants  $K,M >0$ and $\gamma \in [0,1[$ such that 
\begin{align}
\abs{(c_i,\eta)}\leq K M^{\abs{\eta}} \left(\abs{\eta}!\right)^{\gamma},\;\; \forall \eta\in X^\ast,\; \forall i = 1,\ldots,\ell\;, \label{eq:GC_condition}
\end{align}
then $F_c$ constitutes a well defined mapping from $B_{\mathfrak p}^m(R)[t_0,$ $t_0+T]$ into $B_{\mathfrak q}^{\ell}(S)[t_0, \, t_0+T]$ for all $R,T >0$~\cite{Irina_thesis,Winter_Arboleda-etal_2015}. The infimum of all the $\gamma \in [0,1[$ such that~\rref{eq:GC_condition} is satisfied for a series $c \in \allseries{\ell}$ is called the \emph{Gevrey order} of the series $c$.
A series $c \in \allseries{\ell}$ obeying the growth condition in~\rref{eq:GC_condition} is called a {\em globally convergent} series. The set of all globally convergent series in $\allseries{\ell}$ is denoted as $\allseriesGC{\ell}$.
A Fliess operator $F_c$ is {\em globally convergent} if and only if there exists no real number $M > 0$ such that $\rho\left(F_c\right) < M$. Observe that a noncommutative polynomial $\allpoly$ is a globally convergent series with Gevrey degree $0$. As described above, a series $c \in \allseriesGC{\ell}$ is only a sufficient condition for the corresponding Fliess operator $F_c$ to be globally convergent. Necessary conditions are presented in Subsection~\ref{subsec:Frech}. In the absence of any convergence criterion, \rref{eq:Fliess-operator-defined} only defines an operator in a formal sense.

\subsection{Interconnections of Fliess Operators}
\paragraph*{}
Given Fliess operators $F_c$ and $F_d$, where $c,d\in\allseriesellLC$,
the parallel and product connections satisfy $F_c+F_d=F_{c+d}$ and $F_cF_d=F_{c\shuffle d}$,
respectively \cite{Fliess_81}.
When Fliess operators $F_c$ and $F_d$ with $c\in\allseriesellpriLC$ and
$d\in\allseriesmLC$ are interconnected in a cascade fashion, where $\lvert X^{\prime} \rvert = m + 1$, the composite
system $F_c\circ F_d$ has the Fliess operator representation $F_{c\circ d}$, where
the {\em composition product} \cite{Ferfera_80,Gray_CDC09} of $c$ and $d$
is given by
\begeq \label{eq:c-circ-d}
c\circ d=\sum_{\eta\in X^{\prime\ast}} (c,\eta)\,\psi_d(\eta)(\mbf{1}).
\endeq%
Here $\mbf{1}$ denotes the monomial $1\emptyset$, and $\psi_d$ is the continuous (in the ultrametric sense) algebra homomorphism from $\allseriesXpri{}$ to the algebra of $\re$-linear endomorphisms on $\allseries{}$, $\Endallseries{}$, uniquely specified by
$\psi_d(x_i^{\prime}\eta)=\psi_d(x_i^{\prime})\circ \psi_d(\eta)$ with
$ 
\psi_d(x_i^{\prime})(e)= x_0(d_i\shuffle e),
$
$i=0,1,\ldots,m$
for any $e\in\allseries{}$,
and where $d_i$ is the $i$-th component series of $d$
($d_0:=\mbf{1}$). By definition,
$\psi_d(\emptyset)$ is the identity map on $\allseries{}$. The composition product is linear in its left argument.
\begth If $c,c^{\prime} \in \allseriesXpri{\ell}$ and $d \in \allseries{m}$, then $\left(c+c^{\prime}\right) \circ d = c \circ d + c^{\prime} \circ d$.
\endth
When two Fliess operators $F_c$ and $F_d$ are interconnected to form a feedback system with $F_c$ in the forward path and $F_d$ in the feedback path, the generating series of the closed-loop system is denoted by the {\em feedback product} $c@d$. It can be computed explicitly using the Hopf algebra of coordinate functions associated with the underlying {\em output feedback group} \cite{Gray-etal_SCL14}. Define the set of {\em unital} Fliess operators as
$
\Fliessdelta=\{I+F_c\;:\;c\in\allseriesellLC\},
$
where $I$ denotes the identity map. It is convenient to introduce the symbol
$\delta$ as the (fictitious) generating series for the identity map. That is, $F_\delta:= I$ such that 
$I+F_c:=F_{\delta+c}=F_{c_\delta}$ with $c_\delta:=\delta+c$. The set of all such generating series for
$\Fliessdelta$ will be denoted by $\delta  + \allseriesellLC$. The central idea is that $\left(\Fliessdelta,\circ,I\right)$ forms a group of operators under the composition product.
\begdi
F_{c_\delta}\circ F_{d_\delta}=(I+F_c)\circ(I+F_d)
= F_{c_\delta\circ d_\delta},
\enddi
where $c_\delta\circ d_\delta:=\delta+c\circledcirc d$, $c\circledcirc d:=d+c\modcomp d_\delta$, and
$\modcomp$ denotes the {\em mixed} composition product \cite{Gray-Li_05}.
The {\em mixed} composition product definition is induced by the identity $F_{c\modcomp d_\delta}=F_c\circ F_{d_\delta}$ so that
\begdi
c\modcomp d_\delta =\sum_{\eta\in X^\ast} (c,\eta)\,\phi_d(\eta)(\mbf{1}),
\enddi
where $c \in \allseriesXpri{\ell}, d_{\delta} \in \allseriesdelta{m}$ with $\abs{X^{\prime}} = \abs{X} = m+1$, and $\phi_d$ is analogous to $\psi_d$ in \rref{eq:c-circ-d}
except here $\phi_d(x_i)(e)=x_ie+x_0(d_i\shuffle e)$ with $d_0:=0$. Equivalently, $(\allseriesdelta{m},\circ,\delta)$ forms a group. The mixed composition product is also linear in its left argument. The following theorem states that the mixed composition can be viewed as a right group action of $(\allseriesdelta{m},\circ,\delta)$ on $\allseriesXpri{\ell}$.

\begth\label{thm:modcomp_grp} \cite{Gray_Duffaut_CDC_13} If $c \in \allseriesXpri{\ell}$ and $d_\delta,e_\delta \in \allseriesdelta{m}$, then $(c \modcomp d_{\delta})\modcomp e_{\delta} = c \modcomp (d_{\delta} \circ e_{\delta})$.
\endth

The next lemma states that the mixed composition product distributes on the left over the shuffle product.

\begle \cite{Gray-Li_05} \label{le:mixed-comp-dist-shuffle}
If $c,d \in \allseries{\ell}$ with $e \in \allseriesXpri{m}$ such that $\lvert X \rvert = m+1$, then
\begdi \label{lem:shuffle_dist}
(c \shuffle d) \modcomp e_{\delta} = (c \modcomp e_{\delta}) \shuffle (d \modcomp e_{\delta}).
\enddi
\endle

For the group of unital Fliess operators, the coordinate maps for the corresponding Hopf algebra $H$ have the form
\begdi \label{eq:character-maps}
a_\eta:\allseries{\ell}\rightarrow \re^{\ell}:c\mapsto (c,\eta),
\enddi
where $c \in \allseries{\ell}$, $\eta\in X^\ast$.
The commutative product is taken to be the Hadamard product in $\re^\ell$,
\begdi \label{eq:mu-product}
\HAprod:a_\eta\otimes a_{\xi}\mapsto a_{\eta} \odot a_{\xi},
\enddi
where the unit $\mbf{1}$ is defined to map every $c$ to $\doubleone = \left[ 1 1 \cdots 1\right] \in \re^{\ell}$.  If the {\em degree} of $a_{\eta}$ is defined as $\deg(a_{\eta})=2\abs{\eta}_{x_0}+\abs{\eta}_{x_1}+1$, then $H$ is a graded and connected $\re$-algebra with $H=\bigoplus_{k\geq 0} H_k$, where $H_k$ is the set of all elements of degree $k$ and $H_0=\re\mbf{1}$ \cite{Foissy_15}. The coproduct $\Delta$ is defined so that the formal power series product
$c\circledcirc d$ for the group $\allseriesdelta{\ell}$ satisfies
\begdi
\Delta a_\eta(c,d)=a_\eta(c\circledcirc d)=(c\circledcirc d,\eta).
\enddi
Of primary importance is the following lemma which describes how
the group inverse $c_\delta^{\circ -1}:=\delta+c^{\circ -1}$ is computed.

\begle \cite{Gray-etal_SCL14} \label{le:antipode-is-group-inverse}
The Hopf algebra $(H,\HAprod,\Delta)$ has an antipode $S$ satisfying
$a_\eta(c^{\circ -1})=(Sa_\eta)(c)$ for all $\eta\in X^\ast$ and $c\in\allseries{\ell}$.
\endle

With this concept, the generating series for the feedback connection, $c@d$, can be computed
explicitly.
\begth \cite{Gray-etal_SCL14}
\label{th:feedback-product-formula}
For any $c\in\allseries{\ell}$ and $d \in \allseriesXpri{m}$, where $\lvert X \rvert = m +1$ and $\lvert X^{\prime} \rvert = \ell + 1$, it follows that
$
c@d=c\modcomp(-d\circ c)_\delta^{\circ -1}.
$
\endth

\subsection{Fr\'echet Topology for Global Convergence}\label{subsec:Frech}
\paragraph*{}
The ultrametric topology on $\allseries{\ell}$ provides a framework to prove the existence of a well-defined feedback product via fixed point theorems as described in Section \ref{sec:static-feedback}. However, a convergent sequence of series in the ultrametric space $\allseries{\ell}$, each of which has a well-defined Fliess operator, need not have a well-defined Fliess operator corresponding to the limit. This is demonstrated by the following example.

\begex\label{ex:frech} Let $\left(c_i\right)_{i \in \nat_0}$ be a sequence of series in $\allseries{}$. Let 
\begin{align*}
c_i = \sum_{k = 0}^i \left(k!\right)^{1+\epsilon} x_1^k,  
\end{align*}
where $\epsilon > 0$. Observe that each $c_i$ is a polynomial; hence, $c_i \in \allseriesGC{}$. It is evident that the sequence $\left(c_i\right)_{i \in \nat_0}$ is Cauchy in the ultrametric topology. The sequence $c_i \longrightarrow c$, where $c$ is defined as
\begin{align*}
c = \sum_{k = 0}^\infty \left(k!\right)^{1+\epsilon} x_1^k.
\end{align*}  
Since $\epsilon > 0$, there exist no constants $K,M >0 $ such that $\abs{\left(c,x_1^n\right)} \leq KM^{n}n! \; \forall n \in \nat_0$. Therefore, $c \, \notin \,\allseriesLC{}$. 
\endex

This subsection describes the construction of a topology called the \emph{Fr\'echet} or \emph{seminorm} topology under which global convergence of Fliess operators is preserved in the limit. 

\begde\label{def:norm_conv} Let $c \in \allseries{}$. Then, for any positive real number $R > 0$, define the map $\norminf{R}{.} : \allseries{} \mapsto \overline{\re_{+}}$ as
\begin{align*}
\norminf{R}{c} &= \sup\limits_{\eta \in \allwords}\left\lbrace\abs{\left(c,\eta\right)}\frac{R^{\abs{\eta}}}{\abs{\eta}!}\right\rbrace,
\end{align*}
where $\overline{\re_{+}}$ is the closure of the non-negative real line with $+\infty$. For all positive real $R >0$, define the normed space
\begin{align*}
\serSinfR{m}{R} &= \left\lbrace c \in \allseries{m} \; :\norminf{R}{c_i} < \infty \; \forall i=1,\ldots,m\right\rbrace.
\end{align*}
\endde

The superscript is omitted when $m=1$. Note that $\serSinfR{}{R}$ is isometrically isomorphic to the Banach space $\ell^{\infty}\left(X^{\ast}\right)$, the space of all bounded functions from $X^{\ast}$ to $\re$. Hence, the tuple $\left(\serSinfR{}{R},+,\cdot,\norminf{R}{.}\right)$ forms an infinite dimensional Banach space, where $+$ and $\cdot$ represent series addition and scalar multiplication, respectively. The following theorem states that a formal power series $c$ is {\em locally convergent} as in~\rref{eq:Gevrey-growth-condition} if and only if $c$ belongs to $\serSinfR{}{R}$ for some $R > 0$.

\begth\cite{Irina_thesis} $\allseriesLC{} = \bigcup_{R > 0} \serSinfR{}{R}$. 
\endth

The locally convex space $\serSinfR{}{R^{\prime}}$ is an infinite dimensional Banach space in which the standard Bolzano-Wierstrass theorem fails to hold. Hence, not every bounded sequence in $\serSinfR{}{R^{\prime}}$ has a convergent subsequence in $\serSinfR{}{R^{\prime}}$ as shown in the following example. 

\begex Consider the sequence of the series $\left(c_i\right)_{i \in \nat_0} \in \serSinfR{}{R}$ such that 
\begin{align*} 
c_n = \sum_{\eta \in X^n} \left(\frac{1}{R}\right)^n n!\,\eta.
\end{align*}
It is evident that the sequence $\left(c_i\right)_{i \in \nat_0}$ is bounded as $\norminf{R}{c_n} = 1 \; \forall n \in \nat_0$. However, note that $\forall \; m,n \in \nat_0$ where $m \neq n$,
\begin{align*}
\norminf{R}{c_m -c_n} = 1.
\end{align*}
Hence, the bounded sequence $\left(c_i\right)_{i \in \nat_0}$ has no convergent subsequence. 
\endex

Moreover, the space $\serSinfR{}{R^{\prime}}$ is not a separable space, viz. the Banach space does not have a countable dense topological subspace \cite{Dahmen-etal_2020}. The space $\serSinfR{m}{R^{\prime}}$, which is a direct product of $m$ Banach spaces, is provided a Banach space structure by the norm
\begin{align*} 
\norminf{R}{d} &= \max_{i = 1,2,\ldots,m} \norminf{R}{d_i}.
\end{align*}

Let $R^{\prime},R \in \re$ such that $0 < R < R^{\prime}$. Observe that $\serSinfR{}{R^{\prime}} \subset \serSinfR{}{R}$ as vector spaces. In addition, the topology on $\serSinfR{}{R^{\prime}}$ induced by the norm $\norminf{R^{\prime}}{.}$ is finer than the subspace topology induced from $\serSinfR{}{R}$. Hence, this inclusion of vector spaces is not a topological embedding. In fact, the inclusion map $i : \serSinfR{}{R^{\prime}} \longrightarrow \serSinfR{}{R}$ is a compact operator, viz. every bounded sequence in $\serSinfR{}{R^{\prime}}$ has a convergent subsequence in $\serSinfR{}{R}$ \cite{Dahmen-etal_2020}. 

Consider the directed set $\left(\re_{> 0}, \leq\right)$ with the usual ordering. Then $S_{*} = \{\left(\serSinfR{}{R}\right)_{R \in \re_{> 0}}\}$ forms a projective system of locally convex topological vector spaces with the family of inclusion maps $i_{R^{\prime}R} : \serSinfR{}{R^{\prime}} \longrightarrow \serSinfR{}{R} \; \forall 0 < R < R^{\prime}$. The projective limit of the system $\left(S_{*},\left(i_{R^{\prime}R}\right)_{0<R<R^{\prime}}\right)$ is a locally convex topological vector space $\serSinf{}$ defined as
\begin{align*}
\serSinf{} &= \bigcap_{R \in \re_{> 0}} \serSinfR{}{R}.
\end{align*}
The limit space $\serSinf{}$ is equipped with the initial topology determined by the family of canonical injections $i_R : \serSinf{} \longrightarrow \serSinfR{}{R} \; \forall R >0$. Thus,
\begin{align*}
c \in \serSinf{} \Leftrightarrow \norminf{R}{c} < \infty \; \forall R > 0.
\end{align*} 
Since the set $\nat \subset \re_{> 0}$ is cofinal, it is sufficient to consider the space $\serSinf{}$ as the projective limit of the spaces $\serSinfR{}{N}$, where $N \in \nat$. Hence, the space $\serSinf{}$ is the sequential projective limit of the Banach spaces $\left(\serSinfR{}{N}\right)_{N \in \nat}$ and can be endowed with the Fr\'echet topology~\cite{Carreras-Bonet_87}. The ordered set of countable seminorms 
\begin{align*}
\norminf{1}{.} \leq \norminf{2}{.} \leq \cdots \leq \norminf{k}{.} \leq \cdots
\end{align*}
is called \emph{a fundamental system of seminorms} for the Fr\'echet space. The Fr\'echet spaces are completely metrizable locally convex topological vector spaces. Hence, 
\begin{align*}
\left(c_i\right)_{i \in \nat_0} \rightarrow c \in \serSinf{} \Leftrightarrow \lim_{i \rightarrow \infty} \norminf{R}{c_i-c} = 0 \quad \forall R > 0.
\end{align*}
Since the inclusion maps $i_{MN}$ where $0< N < M$ with $M,N \in \nat$ are compact operators, the projective limit $\serSinf{}$ becomes a \emph{Fr\'echet-Schwartz} space \cite{Carreras-Bonet_87,Komatsu_67}. Thus, the space $\serSinf{}$ is separable. A countable dense topological subspace is constructed in \cite{Irina_thesis}. Hence, the limit space $\serSinf{}$ is better behaved than the spaces $\serSinfR{}{R}$ from which it is constructed. In particular, the space $\serSinf{}$ satisfies a Bolzano-Wierstrass like theorem. The space $\serSinf{m} \subset \allseries{m}$ is defined as 
\begin{align*}
c \in \serSinf{m} \Leftrightarrow c_i \in \serSinf{} \; \forall i =1,\ldots,m
\end{align*}
and is endowed with the \emph{product} topology whenever $m > 1$. The construction of the space $\serSinf{m}$ is pivotal in regards to describing the radii of convergence of Fliess operators. 

\begth\cite{Irina_thesis}\label{thm:finite_radius}
A series $c \in \serSinfR{m}{R}$ for some $R>0$ if and only if the corresponding Fliess operator $F_c$ is locally convergent. 
\endth  
Observe that if $c \in \allseriesGC{}$, then $\norminf{R}{c} < \infty \; \forall R > 0$. Hence, $\allseriesGC{m}\subset\serSinfR{m}{R} \; \forall R >0$ implying that $\allseriesGC{m} \subset \serSinf{m}$.
\begth\cite{Irina_thesis} \label{thm:GC_closure}$\serSinf{m} = \overline{\allseriesGC{m}}$, where the closure is taken in the Fr\'echet topology. 
\endth
\begth\cite{Irina_thesis}\label{thm:infinite_radius} A series $c \in \serSinf{m}$ if and only if the corresponding Fliess operator $F_c$ is globally convergent.
\endth

Theorem~\ref{thm:infinite_radius} asserts that it is necessary and sufficient for a series $c \in \serSinf{m}$ in order for the corresponding Fliess operator $F_c$ to describe a well-defined mapping from $B_{\mathfrak p}^m(R)[t_0,$ $t_0+T]$ into $B_{\mathfrak q}^{\ell}(S)[t_0, \, t_0+T]$ for all $R,T >0$. Observe that in Example~\ref{ex:frech} the sequence of polynomials is not even Cauchy in $\serSinfR{}{R}\;\forall R > 0$. Hence, the sequence does not converge in the $\serSinf{m}$ space. Define $\partial\allseriesGC{m} = \serSinf{m}\setminus\allseriesGC{m}$ in the Fr\'echet topology. The following example shows that the boundary $\partial\allseriesGC{}$ is not empty.

\begex\cite{Irina_thesis}\label{ex:Ferafera_bound} Let $X  = \{x_0,x_1\}$. The Ferfera series $c \in \allseriesGC{}$ is given by $c = x_1^{\ast} = \sum_{n = 0}^\infty x_1^n$. It is evident that $c \in \allseriesGC{}$ with Gevrey degree $0$. Consider the series $d = c \circ c$ which describes the cascade connection of two Ferfera systems. It is known that $d$ has Gevrey degree $1$, but the corresponding Fliess operator $F_d$ has a well-defined mapping from $B_{\mathfrak p}^m(R)[t_0,$ $t_0+T]$ into $B_{\mathfrak q}^{\ell}(S)[t_0, \, t_0+T]$ for all $R,T >0$. Hence, there exists a series $c \in \partial\allseriesGC{m}$ with the Gevrey degree $1$ such that the corresponding Fliess operator has a well-defined mapping globally.  
\endex

\begin{figure}[t]
	\centering
	\includegraphics[scale = 0.75]{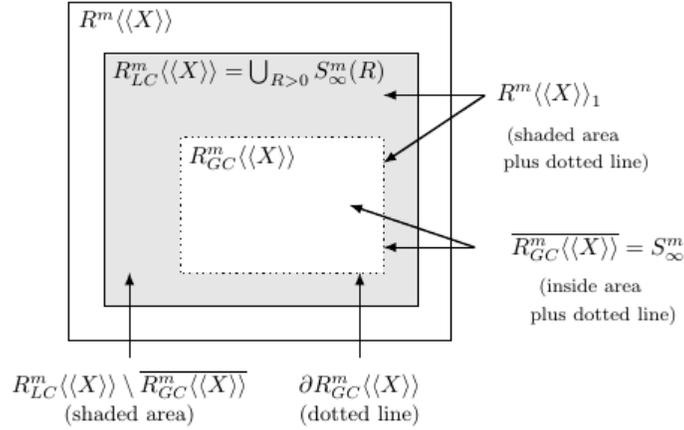}
	\caption{The hierarchy of topological vector spaces for convergence.}
	\label{fig:Class-series-norm-spaces}
\end{figure}
Define $\allseries{m}_1$ as the set of series with Gevrey degree $1$. The hierarchy of the spaces $\allseriesLC{m},\, \serSinf{m},\, \allseriesGC{m}$ and $\allseries{m}_1$ are depicted in  Figure~\ref{fig:Class-series-norm-spaces}. The following theorem describes the algebraic closure of local and global convergence of series under addition.
\begth\cite{Irina_thesis}\label{thm:addition_conv} The following statements are true:
\begin{enumerate}
	\item If $c,d \in \allseriesLC{m}$, then $c+d \in \allseriesLC{m}$.
	\item If $c,d \in \allseriesGC{m}$, then $c+d \in \allseriesGC{m}$.
	\item If $c,d \in \serSinf{m}$, then $c+d \in \serSinf{m}$.
\end{enumerate}
\endth 

The following results are necessary for proving the global convergence of the Wiener-Fliess composition product, which is presented in Section~\ref{sec:glo_conv}. Theorem~\ref{thm:shuffle_GC} states that the shuffle product is closed in $\allseriesGC{}$.

\begth\label{thm:shuffle_GC}\cite{GS_thesis} Let $c_1,c_2,\ldots,c_k$ be a finite nonempty collection of formal power series with $c_i \in \allseriesGC{} \;\forall i = 1,2,\ldots,k$. Then, $c_1\shuffle c_2\shuffle\cdots\shuffle c_k \in \allseriesGC{}$.
\endth

The following corollary is an immediate result of Theorem~\ref{thm:shuffle_GC}.
\begco\label{cor:shuffle_GC_pow}\cite{GS_thesis} If $c \in \allseriesGC{}$, then $c^{\shuffle n} \in \allseriesGC{} \; \forall n \in \nat_0$.
\endco 

The following theorem characterizes a kind of {\em almost} submultiplicative property of $\norminf{R}{\cdot}$ with respect to the shuffle product of formal power series.  
\begth\label{thm:shuffle_global_lem}\cite{GS_thesis} Let $c_1,c_2,\ldots,c_k$ be a finite nonempty collection of formal power series with $c_i \in \serSinfR{}{R} \;\forall i = 1,2,\ldots,k$. Then, $c_1\shuffle c_2\shuffle\cdots\shuffle c_k \in \serSinfR{}{R^{\prime}} \; \forall R^{\prime} = \epsilon R$, where $\epsilon \in ]0,1[$ and
\begin{align*}
\norminf{R^{\prime}}{c_1\shuffle c_2 \shuffle\cdots\shuffle c_k} &\leq \frac{1}{(1-\epsilon)^k}\norminf{R}{c_1}\norminf{R}{c_2}\cdots\norminf{R}{c_k}.
\end{align*}
\endth 

The following corollary is a consequence of Theorem~\ref{thm:shuffle_global_lem} and is used in Section~\ref{sec:glo_conv}.
\begco\label{cor:shuffle_power_conv}\cite{GS_thesis} If $c \in \serSinfR{}{R}$, then $\forall R^{\prime}=\epsilon R$, where $\epsilon \in ]0,1[$, it follows that
\begin{align*}
\norminf{R^{\prime}}{c^{\shuffle k}} \leq \frac{\norminf{R}{c}^k}{(1-\epsilon)^k}.
\end{align*} 
\endco

\subsection{Formal Static Maps and Convergence}
\paragraph*{}
This subsection provides a brief discussion on formal static maps. Let $\tilde{X} = \{\tilde{x}_1,\ldots,\tilde{x}_m\}$ and $d \in \allcommutingseriesX{k}$. A formal static function $f_d : \re^m \longrightarrow \re^k$  around the point $z = 0$ is defined as 
\begin{align*}
f_d\left(z\right) &= \sum_{\eta \in \tilde{X}^{\ast}} \left(d,\eta\right) z^{\eta},
\end{align*}
where $z \in \re^m$, and $z^{\tilde{x}_i\eta}= z_iz^{\eta} \; \forall \tilde{x}_i \in \tilde{X}, \eta\in \allcommutingwords$. The base case is taken to be $z^{\emptyset} = 1$. The series $d \in \allcommutingseriesX{k}$ is called the generating series of the static map $f_d$. A series $d \in \allcommutingseriesX{}$ is said to be \emph{locally convergent} if there exist constants $K_d,M_d >0$ such that $\abs{\left(d,\eta\right)} \leq K_dM_d^{\abs{\eta}}, \; \forall \eta \in \tilde{X}^{\ast}$. A series $d \in \allcommutingseriesX{k}$ is said to be locally convergent if and only if each component $d_i$ is locally convergent for $i=1,\ldots, m$. The subset of all locally convergent series in $\allcommutingseriesX{k}$ is denoted as $\allcommutingseriesLCX{k}$. The following theorem describes the significance of the definition of local convergence in the present context.

\begth If $d \in \allcommutingseriesLCX{}$ with growth constants $K_d,M_d >0$ where $M_d = \inf\{M : \abs{\left(d,\eta\right)} \leq K_dM^{\abs{\eta}} \; \forall \eta \in \allcommutingwords \}$, then the formal static function $f_d : \re^m \longrightarrow \re$ has a finite radius of convergence $1/M_d$. 
\endth

\begpr Let $z \in \re^m$. From the triangle inequality on $\re$,
\begin{align*}
\abs{f_d\left(z\right)} &\leq \sum_{\eta \in \tilde{X}^{\ast}} \abs{\left(d,\eta\right)} \abs{z^\eta} \\
&\leq \sum_{\eta \in \tilde{X}^{\ast}} K_dM_d^{\abs{\eta}} \abs{z^{\eta}} \\
&= K_d \left(\sum_{n = 0}^\infty \left(M_d\abs{z^{\tilde{x}_1}}\right)^n\right)\cdots \left(\sum_{n = 0}^\infty \left(M_d\abs{z^{\tilde{x}_m}}\right)^n\right) \\
&= K_d \prod_{i=1}^m  \left(\sum_{n = 0}^\infty \left(M_d\abs{z_i}\right)^n\right).
\end{align*}
Observe that
\begin{align*}
\sum_{n = 0}^\infty \left(M_d\abs{z_i}\right)^n = \left(\frac{1}{1-M_d\abs{z_i}}\right)
\end{align*}
for $\abs{z_i} \leq 1/M_d$. Hence,
\begin{align*}
\abs{f_d\left(z\right)} &\leq K_d \prod_{i=1}^m \left(\frac{1}{1-M_d\abs{z_i}}\right)
\end{align*} 
for $\max\limits_{i=1,\ldots,m} \abs{z_i} \leq 1/M_d$.
\endpr 

Therefore, $d \in \allcommutingseriesLCX{}$ is sufficient for the corresponding static function $f_d$ is bounded pointwise in absolute value by a real analytic map with a finite radius of convergence. However, the condition $d \in \allcommutingseriesLCX{}$ is indeed necessary and follows from the Cauchy's Integral Formula on polydiscs in $\mathbb{C}^m$\cite{Hormander_03}. Hence, $d \in \allcommutingseriesLCX{}$ is a necessary and sufficient condition for the corresponding static map $f_d$ to be locally analytic around $z = 0$. The following lemma is essential in proving the sufficient condition for global convergence of static maps and for further results in Section~\ref{sec:glo_conv}.
\begle\label{lem:factorial}\cite{GS_thesis} Given $x \in [0,\infty[$ and $r \in ]0,1]$, the following inequality holds:
\begin{align*}
K_rM_r^x (\Gamma(x+1))^r \leq \Gamma(rx+1) \leq \tilde{K}_r2^x(\Gamma(x+1))^r,
\end{align*} 
where $\Gamma\left(\cdot\right)$ denotes the Gamma function, and 
\begdi 
K_r = \left(\left(\frac{2\pi}{\exp(2)}\right)^{1-r}r\right)^{\frac{1}{2}}, \; \;
\tilde{K}_r = 2\left(\left(\frac{2\pi}{\exp(2)}\right)^{1-r}4\right)^{\frac{1}{2}}, \;\;
M_r = r^r.
\enddi
\endle

The \emph{Gevrey order} of a series $d \in \allcommutingseries{}$ is defined as
\begin{align*}
s = \inf\{t \geq 0 : \abs{\left(d,\eta\right)} &\leq K_dM_d^{\abs{\eta}}\left(\abs{\eta}!\right)^{t}, \; \forall \eta \in \allcommutingwords, K_d , M_d \in \re_{+} \}.
\end{align*}
A series $d \in \allcommutingseries{}$ is said to be \emph{globally convergent} if there exist constants $K_d, M_d >0$ and $s \in \left[0,1\right[$ such that 
\begin{align*}
\abs{\left(d,\eta\right)} &\leq K_dM_d^{\abs{\eta}}\left(\abs{\eta}!\right)^{-1+s}, \quad \; \forall \eta \in \allcommutingwords.
\end{align*}
Hence, a series $d \in \allcommutingseriesGCX{}$ has a Gevrey order $\left(-1+s\right)$ with $s \in [0,1[$ while a series $d \in \allcommutingseriesLCX{}$ has a Gevrey order of $0$. A series $d \in \allcommutingseriesX{k}$ is said to be globally convergent if and only if each component $d_i$ is globally convergent for $i=1,\ldots,m$. The subset of all globally convergent series in $\allcommutingseriesX{k}$ is denoted as $\allcommutingseriesGCX{k}$. The following theorem explains the significance of the definition of global convergence of a series with respect to its corresponding static function.

\begth\label{thm:formal_static_GC} If $d \in \allcommutingseriesGCX{}$ with growth constants $K_d,M_d >0$ and Gevrey order $\left(-1+s\right)$ with $s \in [0,1[$, then the formal static function $f_d : \re^m \longrightarrow \re$ converges over the entire domain $\re^m$.
\endth

\begpr Let $z \in \re^m$. From the triangle inequality on $\re$,
\begin{align*}
\abs{f_d\left(z\right)} &\leq \sum_{\eta \in \tilde{X}^{\ast}} \abs{\left(d,\eta\right)} \abs{z^{\eta}} \\
&\leq \sum_{\eta \in \tilde{X}^{\ast}} K_dM_d^{\abs{\eta}}\left(\abs{\eta}!\right)^{-1+s} \abs{z^{\eta}} \\
&\leq K_d \left(\sum_{n = 0}^\infty \frac{\left(M_d\abs{z^{\tilde{x}_1}}\right)^n}{n!^{\left(1-s\right)}}\right)\cdots \left(\sum_{n = 0}^\infty \frac{\left(M_d\abs{z^{\tilde{x}_m}}\right)^n}{n!^{\left(1-s\right)}}\right) \\
&= K_d \prod_{i=1}^m  \left(\sum_{n = 0}^\infty \frac{\left(M_d\abs{z_i}\right)^n}{n!^{\left(1-s\right)}}\right).
\end{align*}
Since $n! = \Gamma\left(n+1\right)$, by Lemma~\ref{lem:factorial},
\begin{align*}
\left(\sum_{n = 0}^\infty \frac{\left(M_d\abs{z_i}\right)^n}{n!^{\left(1-s\right)}}\right) &\leq \tilde{K}_r \left(\sum_{n = 0}^\infty \frac{\left(2M_d\abs{z_i}\right)^n}{\Gamma\left(\left(1-s\right)n+1\right)}\right) \\
&= \mathbb{E}_{\left(1-s\right),1}\left(2M_d\abs{z_i}\right),
\end{align*}
where $\mathbb{E}_{\left(1-s\right),1}\left(.\right)$ is the Mittag-Leffler function. Hence,
\begin{align*}
\abs{f_d\left(z\right)} &\leq K_d \prod_{i=1}^m \mathbb{E}_{\left(1-s\right),1}\left(2M_d\abs{z_i}\right).
\end{align*}
Observe that $s \in [0,1[$ if and only if $\left(1-s\right) \in ]0,1]$. Hence, $\mathbb{E}_{\left(1-s\right),1}\left(\cdot\right)$ is an entire function on $\mathbb{C}$. Therefore, $d \in \allcommutingseriesGCX{}$ implies that the corresponding static map $f_d$ is bounded pointwise in absolute value by a real analytic map which is convergent everywhere on $\re^m$. 
\endpr

Observe that a commutative polynomial $d \in \allcommutingpolyX{}$ is globally convergent with Gevrey order $-1$. As in the case of Chen-Fliess series, a commutative series $d \in \allcommutingseriesGCX{m}$ is only a sufficient condition for the corresponding formal static map to be convergent everywhere on $\re^m$. The derivation of a necessary condition for a real analytic function that is convergent everywhere on $\re^m$ requires more careful attention. A function that is real analytic everywhere on $\re^m$ need not extend to an entire function upon complexification of the domain.  
\begex Consider $f : \re \longrightarrow \re$ defined as $f(x)= 1/\left(x^2 +1\right)$. The function $f$ is analytic everywhere on $\re$. The complexification of $f : \mathbb{C} \longrightarrow \mathbb{C}$ given by $f(z) = 1/\left(z^2 +1\right)$ is not an entire function on $\mathbb{C}$ as the complex map $f$ has poles at $z = \pm i$. 
\endex

A locally real analytic function always extends to a locally analytic complex function, but a function that is analytic over the entire real line does not necessarily extend to an entire function. Hence, the complexification approach does not yield a necessary growth condition for a real analytic function that is analytic everywhere on $\re^m$. The derivation of a necessary condition is deferred to future work.

\subsection{Shuffle Group}
\paragraph*{}
This subsection presents the \emph{shuffle group}. The computations and algorithms described in  Section~\ref{sec:HA-shuffle-group} are based on the Hopf algebra of the coordinate maps defined on the shuffle group. The following theorem describes the {\em shuffle group}.
\begth \cite{Gray-etal_AUTO14} \label{th:shuffle-group}
The set of non-proper series in $\allseries{}$ is a group under the shuffle
product. In particular, the shuffle inverse of any such series $c$ is
\begdi
c^{\shuffle -1}=((c,\emptyset)(1-c^\prime))^{\shuffle -1}=(c,\emptyset)^{-1}(c^{\prime})^{\shuffle\ast},
\enddi
where $c^\prime:=1-c/(c,\emptyset)$ is proper, $(c^\prime)^{\shuffle\ast}:=\sum_{k\geq 0} (c^\prime)^{\shuffle k}$, and the identity element is the constant $1$.
\endth

More generally, if $c \in \allseries{\ell}$, then the shuffle inverse is defined componentwise, viz. $(c^{\shuffle -1})_{i} = c_i^{\shuffle -1}$, where $i = 1,2,\ldots \ell$. Hence, in general, $(\allseries{\ell},\shuffle)$ possesses a group structure with the identity element $\doubleone \triangleq \left[1\;1 \cdots 1\right]^{T} \in \re^{\ell}$.

\begex\label{ex:shuff_group}
Let $c = 1 - x_1 \; \in \allseries{}$. Observe that, $c' = x_1$, and hence, $c^{\shuffle -1} = x_{1}^{\shuffle \ast} = \sum_{k\geq 0} k!\, x_{1}^k$.
\endex

\section{Wiener-Fliess Connections}\label{sec:Wiener-Fliess}
\begin{figure}[t]
	\begin{center}
		\includegraphics[scale = 0.5]{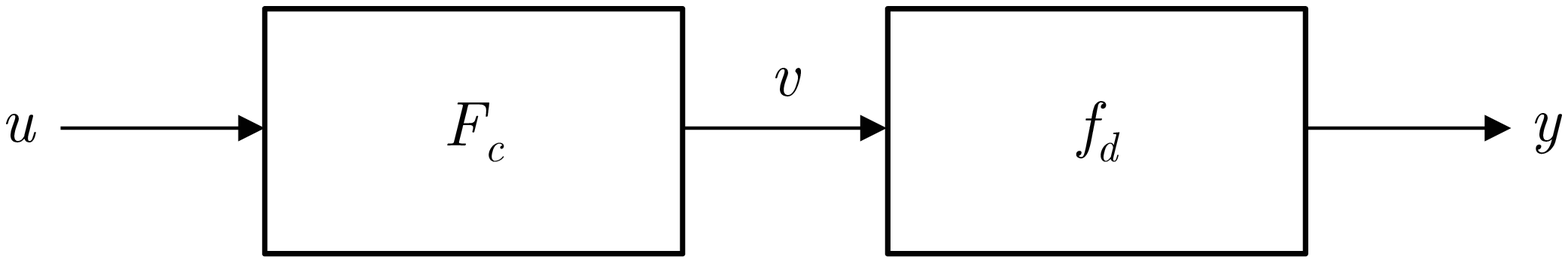}
	\end{center}
	\caption{Wiener-Fliess connection}
	\label{fig:wiener-fliess}
\end{figure}

This section describes the cascade connection shown in Figure~\ref{fig:wiener-fliess} of a Chen-Fliess series $F_c$ generated by a series $c \in \allseriesell$ and a formal static map $f_d : \mathbb{R}^\ell \longrightarrow \mathbb{R}^k $ defined without loss of generality at $z = 0$. Such configurations are called \emph{Wiener-Fliess} connections. The connection is known to generate another well defined formal Fliess operator, and its generating series is computed through the \emph{Wiener-Fliess composition product}. The product is well defined formally due to the local finiteness property in the following cases:
\begin{enumerate}
	\item The Chen-Fliess series $F_c$ is defined by a proper series $c \in \allseriesell$.
	\item The formal static function $f_d : \mathbb{R}^\ell \longrightarrow \mathbb{R}^k $ is a vector of $k$ polynomials. 
\end{enumerate}
The definition addressing the first case appears in \cite{Gray_Thitsa_IJC12}. However, the definition of the \emph{Wiener-Fliess} product remains the same for both cases and is given in the following theorem.

\begth 
\label{thm:Wiener-Fliess-product}
Let $ X = \{x_0,x_1,\ldots,$ $x_m\}$ and $\tilde{X} = \{\tilde{x}_1,\tilde{x}_2,\ldots,\tilde{x}_\ell\}$. Given a formal Fliess operator $F_c$ with $c \in \allseriesell$ and formal function $f_d : \mathbb{R}^\ell \longrightarrow \mathbb{R}^k $ with a generating series $d \in \allcommutingseriesX{k}$ at $z = 0$, viz.
\begin{align*}
f_d(z) =  \sum_{\tilde{\eta} \in \tilde{X}^{\ast}} (d,\tilde{\eta}) z^{\tilde{\eta}},
\end{align*}
the composition $f_d\circ F_c$ has a generating series in $\allseries{k}$ provided either of the following holds:
\begin{enumerate}
	\item $c \in \allseriesell$ is proper.
	\item $d \in \allcommutingpolyX{k}$. 
\end{enumerate}
The generating series of $f_d\circ F_c$ is then given by the Wiener-Fliess composition product
\begin{align}
d \mixcomp c = \sum_{\tilde{\eta} \in \tilde{X}^{\ast}} (d,\tilde{\eta}) c^{\shuffle \tilde{\eta}}\label{eqn:mixcomp},
\end{align}
where $c^{\shuffle\,\tilde{x}_{i}\tilde{\eta}}:=c_{i}\shuffle c^{\shuffle \tilde{\eta}} \; \forall \tilde{x}_i \in \tilde{X} , \; \forall \tilde{\eta} \in \allcommutingwords$, and $c^{\shuffle \phi} = 1$.
\endth

The following theorem shows that the Wiener-Fliess composition product is left $\re$-linear.

\begth If either of the following conditions hold,
\begin{enumerate}
	\item $c \in \allproperseries{\ell}$.
	\item $d,e \in \allcommutingpolyX{k}$.
\end{enumerate}
then $(\alpha d+e)\mixcomp c = \alpha(d \mixcomp c) + (e \mixcomp c)$, where $\alpha \in \re$.
\endth

\begpr 
Observe
\begin{align*}
(\alpha d+e) \mixcomp c &= \sum_{\tilde{\eta} \in \tilde{X}^{\ast}} ((\alpha d)+e,\tilde{\eta}) c^{\shuffle \tilde{\eta}} \\
&= \sum_{\tilde{\eta} \in \tilde{X}^{\ast}} (\alpha d,\tilde{\eta}) c^{\shuffle \tilde{\eta}} + \sum\limits_{\tilde{\eta}
	\in \tilde{X}^{\ast}} (e,\tilde{\eta}) c^{\shuffle \tilde{\eta}} \\
&= \alpha \sum_{\tilde{\eta} \in \tilde{X}^{\ast}} (d,\tilde{\eta}) c^{\shuffle \tilde{\eta}} + \sum_{\tilde{\eta}
	\in \tilde{X}^{\ast}} (e,\tilde{\eta}) c^{\shuffle \tilde{\eta}} \\
&= \alpha (d \mixcomp c) + (e \mixcomp c).
\end{align*}
\endpr

The next lemma will be used to show that the Wiener-Fliess composition product has certain
contractive properties in the ultrametric space $\left(\allseries{\ell},\kappa\right)$.

\begle\label{lem:contract_wf} Let $ X = \{x_0,x_1,\ldots,x_m\}$ and $\tilde{X} = \{\tilde{x}_1,\tilde{x}_2,$ $\ldots,\tilde{x}_\ell\}$. Assume $\eta \in \tilde{X}^{+}$. 
\begin{enumerate}
	\item If $c,\:\tilde{c} \in \allproperseriesell$, then $\kappa(\eta \mixcomp c, \eta \mixcomp \tilde{c}) \leq \; \max \{\kappa(c),\kappa(\tilde{c})\}^{(\lvert \eta \rvert - 1)}\kappa(c,\tilde{c})$.
	\item If $c,\:\tilde{c} \in \allseriesell\setminus\allproperseriesell$, then $\kappa(\eta \mixcomp c, \eta \mixcomp \tilde{c}) \leq \; \kappa(c,\tilde{c})$.
\end{enumerate}
\endle

\vspace*{0.1in}

\begpr
The proof is by induction on the length of $\eta$. If $\eta = \tilde{x}_i$ for $ i = 1,2,\ldots \ell $, then
\begin{flalign*}
\kappa(\tilde{x}_i \mixcomp c, \tilde{x}_i \mixcomp \tilde{c}) &= \kappa(c^{\shuffle \tilde{x}_i},\tilde{c}^{\shuffle \tilde{x}_i})\\
&= \kappa(c_i, \tilde{c}_i) \leq \kappa(c,\tilde{c}).
\end{flalign*}
Hence, the base case is proved. Now assume the hypothesis is true for $ \lvert \eta \rvert  = k \geq 1$. Let $\hat{\eta} = \tilde{x}_j\eta $, where
$\tilde{x}_j \in \tilde{X}$ and $\eta \in \tilde{X}^{k}$. Then
\begin{align*}
\kappa(\hat{\eta} \mixcomp c, \hat{\eta} \mixcomp \tilde{c})
&= \kappa(c^{\shuffle \tilde{x}_j\eta},\tilde{c}^{\shuffle \tilde{x}_j\eta}) \\
&= \kappa(c_j \shuffle c^{\shuffle \eta}, \tilde{c}_j \shuffle \tilde{c}^{\shuffle \eta}) \\
&= \kappa(c_j \shuffle c^{\shuffle \eta} - \tilde{c}_j \shuffle \tilde{c}^{\shuffle \eta})\\
& = \kappa((c_j \shuffle c^{\shuffle \eta} - c_j \shuffle \tilde{c}^{\shuffle \eta}) 
+ (c_j \shuffle \tilde{c}^{\shuffle \eta} - \tilde{c}_j \shuffle \tilde{c}^{\shuffle \eta}))\\
&\leq\max\{ \kappa(c_j \shuffle c^{\shuffle \eta} - c_j \shuffle \tilde{c}^{\shuffle \eta}), 
\kappa(c_j \shuffle \tilde{c}^{\shuffle \eta} - \tilde{c}_j \shuffle \tilde{c}^{\shuffle \eta})\} \\
&=\:\max\{ \kappa(c_j \shuffle (c^{\shuffle \eta} - \tilde{c}^{\shuffle \eta})), 
\kappa((c_j - \tilde{c}_j) \shuffle \tilde{c}^{\shuffle \eta})\}.
\end{align*}
By the triangle inequality on the ultrametric and the induction hypothesis,  
\begin{align*}
\kappa(\hat{\eta} \mixcomp c, \hat{\eta} \mixcomp \tilde{c})
&\leq \; \max\bigg\{ \kappa(c)\:\max\{\kappa(c),\kappa(\tilde{c})\}^{(\lvert \eta \rvert -1)} \kappa(c,\tilde{c}),
\kappa(\tilde{c})^{\lvert \eta \rvert}\:\kappa(c,\tilde{c})\bigg\} \\
& = \; \max\{\kappa(c),\kappa(\tilde{c})\}^{\lvert \eta \rvert} \: \kappa(c,\tilde{c}),
\end{align*}
which proves the claim when $c$ is proper. If $c$ is not proper, then $\kappa(c)=\kappa(\tilde{c})=1$.
Therefore, 
\begin{align*}
\kappa(\hat{\eta} \mixcomp c, \hat{\eta} \mixcomp \tilde{c})
&\leq \; \kappa(c,\tilde{c})
\end{align*}
as desired.
\endpr

For a fixed $d \in \allcommutingseriesX{k}$ define the map $d_{\mixcomp} : \allproperseriesell \longrightarrow \allseries{k}:c \mapsto d\mixcomp c$, and for a fixed $\tilde{d} \in \allcommutingpolyX{k}$ define the map $\tilde{d}_{\mixcomp} : \allseriesell \longrightarrow \allseries{k}:\bar{c} \mapsto d\mixcomp\bar{c}$.
The following theorems describe the contractive properties of $d_{\mixcomp}$ and $\tilde{d}_{\mixcomp}$.

\begth \label{th:Wiener-Fliess-contraction}The map $d_{\mixcomp}$ is a weak contraction map when $\overline{\omega}\:(d) = 1$ and a strong contraction map when $\overline{\omega}\:(d) > 1$.
\endth

\begpr Let $c,c^{\prime} \in \allproperseriesell$. Observe,
\vspace*{-0.01 in}
\begin{align*}
\kappa(d_{\mixcomp}(c),d_{\mixcomp}(c^{\prime})) &= \kappa(d\mixcomp c, d\mixcomp c^{\prime})\\
&= \kappa\left(\sum_{\eta \in \tilde{X}^{\ast}} (d,\eta) (c^{\shuffle \eta} - c^{\prime \shuffle \eta})\right)\\
&\leq \sup\limits_{\eta \in \tilde{X}^{+}} \kappa\big((d,\eta) (c^{\shuffle \eta} - c^{\prime \shuffle \eta})\big) \\
&= \sup\limits_{k \geq \: \overline{\omega}(d)} \sup\limits_{\eta \in \tilde{X}^k} \kappa(c^{\shuffle \eta}, c^{\prime \shuffle \eta}).
\end{align*}
Applying Lemma \ref{lem:contract_wf} gives
\vspace*{-0.01 in}
\begin{align*}
\kappa(d_{\mixcomp}(c),d_{\mixcomp}(c^{\prime})) &\leq \sup\limits_{k \geq \: \overline{\omega}(d)}  \max \{\kappa(c),\kappa(c^{\prime})\}^{k-1} \kappa(c,c^{\prime})\\
&\leq \max \{\kappa(c),\kappa(c^{\prime})\}^{\overline{\omega}(d)-1} \kappa(c,c^{\prime}).
\end{align*}
\endpr

\begth\label{thm:WF_contraction}The map $\tilde{d}_{\mixcomp}$ is a weak contraction map.
\endth

\begpr Let $c,c^{\prime} \in \allseriesell$. Observe,
\vspace*{-0.01 in}
\begin{align*}
\kappa(\tilde{d}_{\mixcomp}(c),\tilde{d}_{\mixcomp}(c^{\prime})) &= \kappa(\tilde{d}\mixcomp c, \tilde{d}\mixcomp c^{\prime})\\
&= \kappa\left(\sum_{\eta \in \supp\left(\tilde{d}\right)} (\tilde{d},\eta) (c^{\shuffle \eta} - c^{\prime \shuffle \eta})\right)\\
&\leq \sup\limits_{\eta \in \supp\left(\tilde{d}\right)} \kappa\big((\tilde{d},\eta) (c^{\shuffle \eta} - c^{\prime \shuffle \eta})\big).
\end{align*} 
By the definition of ultrametric $\kappa$,
\begin{align*}
\kappa(\tilde{d}_{\mixcomp}(c),\tilde{d}_{\mixcomp}(c^{\prime})) &\leq \sup\limits_{\eta \in \supp\left(\tilde{d}\right)} \kappa\left(c^{\shuffle \eta} - c^{\prime \shuffle \eta}\right) \\
&= \sup\limits_{\eta \in \supp\left(\tilde{d}\right)} \kappa\left(c^{\shuffle \eta},c^{\prime \shuffle \eta}\right). 
\end{align*}
Applying Lemma \ref{lem:contract_wf} gives
\begin{align*}
\kappa(\tilde{d}_{\mixcomp}(c),\tilde{d}_{\mixcomp}(c^{\prime})) &\leq \sup\limits_{\eta \in \supp\left(\tilde{d}\right)} \kappa\left(c,c^{\prime}\right) \\
&= \kappa\left(c,c^{\prime}\right).
\end{align*}
\endpr

The following theorem states the associativity property involving the mixed composition product and the Wiener-Fliess composition product. This identity plays a key role in determining the generating series of the static feedback connection in Section~\ref{sec:static-feedback}.
\begth\label{thm:mix-assoc}If either of the following conditions hold,
\begin{enumerate}
	\item $c \in \allproperseries{\ell}$ 
	\item $d \in \allcommutingpolyX{k}$,
\end{enumerate}
with $e \in \allseriesXpri{m}$ such that $\lvert \tilde{X} \rvert = \ell$ and $\lvert X \rvert = m + 1$, then
$d \mixcomp (c \modcomp e_{\delta}) = (d \mixcomp c) \modcomp e_{\delta}$.
\endth

\begpr The proof is obtained directly from
the definition of the Wiener-Fliess composition product in Theorem~\ref{thm:Wiener-Fliess-product}
by linearly extending the identity given in Lemma~\ref{le:mixed-comp-dist-shuffle}.
\endpr

The final theorem of the section states necessary and sufficient conditions for which relative degree is preserved under the Wiener-Fliess composition product.

\begth\label{thm:WF_rel_degree} Let $X = \{x_0,x_1\}$ and $c \in \allproperseries{}$ with relative degree $r_c$. Assume $d \in \allcommutingseriessingvar$. The relative degree of $d\mixcomp c$ is well-defined and equal to $r_c$ if and only if $\left(d,\tilde{x}_1\right) \neq 0$.  
\endth

\begpr The proof follows from the formula in Theorem~\ref{thm:Wiener-Fliess-product} and Lemma~4.6 in \cite{Gray_GS_relative degree}.
\endpr

For the case when $c$ is non-proper and $d \in \allcommutingpolysingvar$, the relative degree of $d\mixcomp c$ requires caution and is hard to characterize in general. The following example demonstrates a case when the Chen-Fliess series $c$ is non-proper but $d \mixcomp c$ has relative degree.  
\begex Let $X = \{x_0,x_1\}$ and $c \in \allseries{}$ such that $c = 1+x_1$. Observe that $c$ has relative degree $1$. Given $d \in \allcommutingpolysingvar$ such that $d = \tilde{x}_1^2$, then
\begin{align*}
d \mixcomp c &= \tilde{x}_1^2 \mixcomp \left(1+x_1\right) \\
&= \left(1+x_1\right)^{\shuffle 2} \\
&= 1 + 2x_1 + 2x_1^2.
\end{align*}
Hence, the relative degree of $d\mixcomp c$ exists and is $1$.
\endex

The following is an example when Chen-Fliess series $c$ is non-proper but $d\mixcomp c$ does not have relative degree.
\begex\label{ex:WF_rel_deg_fail} Let $X = \{x_0,x_1\}$ and $c \in \allseries{}$ such that $c = 1+x_1$. Observe that $c$ has relative degree $1$. Given $d \in \allcommutingpolysingvar$ such that $d = \tilde{x}_1^2 -2\tilde{x}_1$, then
\begin{align*}
d \mixcomp c &= \tilde{x}_1^2-2\tilde{x}_1 \mixcomp \left(1+x_1\right) \\
&= \left(1+x_1\right)^{\shuffle 2} - 2\left(1+x_1\right) \\
&= -1 + 2x_1^2.
\end{align*}
Hence, the relative degree of $d\mixcomp c$ is not well-defined.
\endex

\section{Hopf Algebra of the Shuffle Group}\label{sec:HA-shuffle-group}
\paragraph*{}
The goal of this section is to describe the Hopf algebra of the shuffle group as defined in
Theorem~\ref{th:shuffle-group}. It is utilized subsequently to develop an algorithm to compute the Wiener-Fliess composition product. Define the set of formal power series
\begin{align*}
M &= \{ \doubleone + d \: : d \in \allproperseries{n}\},
\end{align*}
where $\doubleone = \left[ 1 \cdots 1 \: 1 \right]^{T}\in \re^n$. In light of Theorem~\ref{th:shuffle-group}, $(M, \shuffle)$ forms an Abelian group, where the shuffle inverse of $c \in M$ is defined componentwise viz. $(c^{\shuffle -1})_i = (c_{i})^{\shuffle -1}$.  The identity element of the group $M$ is $\doubleone$. Let the set of all maps from $M$ to $\re^n$ be denoted as $\text{Hom}_{\textbf{set}}(M,\re^n)$. The subset $H \subset \text{Hom}_{\textbf{set}}(M,\re^n)$ of coordinate maps defined on group $M$ is
\begin{align*}
H &= \{ a_{\eta} \: : a_{\eta}(c) = (c,\eta) \: : \: \eta \in X^{*} \}.
\end{align*}
$H$ has an $\mathbb{R}$-algebra structure with addition, scalar multiplication and product defined, respectively, as
\begin{align*}
(a_{\eta} + a_{\zeta})(c) &= a_{\eta}(c) + a_{\zeta}(c)\\
(k a_{\eta})(c) &= k (a_{\eta}(c)) \\
\HAprod(a_{\eta},a_{\zeta})(c) &= a_{\eta}(c)\odot a_{\zeta}(c),
\end{align*}
where $\eta,\zeta \in X^{*}, k \in \mathbb{R}$, and $\odot$ denotes the Hadamard product on $\re^n$.
The unit for the product is given by $a_{\emptyset}$ with $a_{\emptyset} (c) = \doubleone$, $\forall c \in M$.
Define the coproduct $\Delta : H \longrightarrow H \bigotimes H$ as $\Delta a_{\eta}(c,d) = a_{\eta}(c \shuffle d)$,
where $c,d \in M$ and $\eta \in X^{*}$.
The counit map $\epsilon$ is defined as
\begin{align*}
\epsilon(a_{\eta}) = \begin{cases}
1\; : & \!\!\!\eta = \emptyset \\
0\; : & \!\!\!\text{otherwise}.
\end{cases}
\end{align*}
It is simple to check that $(H,\HAprod,a_{\emptyset},\Delta,\epsilon)$ forms a commutative and cocommutative unital bialgebra. The bialgebra is graded based on word length viz. $H = \bigoplus_{k \in \mathbb{N}_0} H_k$ with $a_{\eta} \in H_k$ if and only if $\lvert \eta \rvert = k$. Since $\re \cong H_0$ in the category of algebras with $\epsilon$ acting as the isomorphism, $H$ is a connected and graded bialgebra. The reduced coproduct $\Delta^{\prime}$ is defined as
$\Delta^{\prime}(a_{\eta}) = \Delta(a_{\eta}) - a_{\eta} \otimes \doubleone - \doubleone \otimes a_{\eta}$ if $\eta \neq \emptyset$. Here, by abuse of notation, $\doubleone$ also stands for the constant map which takes $c$ to $\doubleone$ for all $c \in M$. For the case of the empty word, $\Delta^{\prime}(a_{\emptyset}) = 0$. If $c,d \in \allproperseries{n}$, then their corresponding elements in the shuffle group $M$ are $\doubleone + c$ and $\doubleone + d$, respectively. The shuffle product of two proper series is computed by the reduced coproduct of the corresponding elements in the shuffle group $M$.
For all proper series $c,d \in \allproperseries{n}$ and $\eta \in \allwords$, it
follows that $(c\shuffle d,\eta) = \Delta^{\prime}(a_{\eta})(\doubleone + c, \doubleone + d)$.
The antipode map $S : H \longrightarrow H$ is given by $S(a_{\eta})(c) = a_{\eta}(c^{\shuffle -1})$. Since the
Hopf algebra is graded and connected the antipode can be computed for any $a \in H^+$ (where $H^+ := \bigoplus_{k \geq 1}H_k$ ) as \cite{Figueroa-Gracia-Bondia_05}
\begin{align*}
S(a) &= -a - \sum a^{\prime}_{(1)}\odot S(a^{\prime}_{(2)}),
\end{align*}
where the summation is taken over all components of the reduced coproduct 
$\Delta^{\prime}(a)$ written in the Sweedler notation \cite{Sweedler,Abe_Hopf}. Therefore,
the tuple $(H,\HAprod,a_{\emptyset},\Delta,\epsilon,S)$ forms a commutative, cocommutative, connected and graded unital Hopf algebra.

\begex Reconsider Example~\ref{ex:shuff_group}, where $c = 1-x_1 \in \allseries{}$ so that $c^{\shuffle -1} = \sum_{k \geq 0} k!\,x_1^{k} $. The goal is to determine $\left(c^{\shuffle -1}, x_1^2\right)$ directly without computing the entire shuffle inverse. Observe
\begin{flalign*}
a_{x_1^2} (c^{\shuffle -1}) &= S(a_{x_1^2}) (c),
\end{flalign*}
and the reduced coproduct of $a_{x_1^2}$ is
\begin{flalign*}
\Delta'(a_{x_1^2}) &= 2 (a_{x_1} \otimes a_{x_1}).
\end{flalign*}
Since $\Delta'(a_{x_1}) = 0$, it follows that
\begin{flalign*}
S(a_{x_1}) &= -a_{x_1}.
\end{flalign*}
Hence,
\begin{flalign*}
S(a_{x_1^2}) (c) &= -a_{x_1^2}(c) - 2 \big(a_{x_1}(c)(-a_{x_1}(c))\big)\\
&= 0 -2(1(-1)) = 2.
\end{flalign*}
Therefore, $(c^{\shuffle -1},x_1^2) = 2$, as verified by applying Theorem~\ref{th:shuffle-group}.
\endex 

An inductive algorithm is presented next to compute the coproduct $\Delta$ on $H$. A key feature of the algorithm is a
recursively defined \textit{partition} map $\mu : \: X^{\ast} \longrightarrow X^{\ast} \otimes X^{\ast}$,
where $x_{j}\eta \: \mapsto (x_j \otimes \emptyset \:+\: \emptyset \otimes x_j)\mu(\eta)$ with $\eta \in X^{\ast}$, $x_j \in X$, and $\mu(\emptyset) := (\emptyset \otimes \emptyset)$. The definition of the map $\mu$ is exactly dual to the definition of the \textit{deshuffle} coproduct $\Delta_{\shuffle}$ described in \cite{Foissy_15}. The deshuffle coproduct is described on the coordinate maps $a_{\eta}$ for all $\eta\in \allwords$ and involves the splitting of the coordinate maps. However, from an algorithmic perspective, it is more natural to split the underlying words as described in the following algorithm.
\begalg\label{alg:cop}
For all $\eta \in X^{\ast}$ and $c,d \in M$, the coproduct $\Delta a_{\eta}(c,d)$ can be computed as:
\begin{enumerate}
	\item $\mu(\eta) = \sum \eta_{(1)} \otimes \eta_{(2)}$.
	\vspace*{0.05 in}
	\item $\Delta a_{\eta}(c,d) = \sum a_{\eta_{(1)}}(c) \odot a_{\eta_{(2)}}(d)$.
\end{enumerate}
\endalg

This algorithm can be directly extended to compute the reduced coproduct.
\begalg\label{alg:redcop}
For all $\eta \in X^{\ast}$ and $c,d \in M$, the reduced coproduct $\Delta^\prime a_{\eta}(c,d)$ can be computed as:
\begin{enumerate}
	\item If $\eta = \emptyset$, then $\Delta^{\prime}a_{\eta}(c,d) = 0$.
	\item Else, $\Delta^{\prime}a_{\eta}(c,d) = \Delta a_{\eta}(c,d) - a_{\eta}(c) \odot \doubleone - \doubleone \odot a_{\eta}(d)$.
\end{enumerate}
\endalg

Let $\Phi_{c}$ be an $\re$-linear homomorphism of algebras defined as  $\Phi_{c} : H \longrightarrow \re^n : a_{\eta} \mapsto a_{\eta}(c)$, where $\re^n$ is an $\re$-algebra under the Hadamard product. The maps $\Phi_{c}$ are usually called the \textit{characters} of the Hopf algebra $H$ and form a group under the Hopf convolution product $\star$ defined as
\begin{align*}
(\Phi_{c}\star\Phi_{d}) (a_{\eta}) &= \HAprod \circ (\Phi_c \otimes \Phi_d) \circ \Delta(a_{\eta})  \\
&= \sum \Phi_c(a_{\eta_{(1)}}) \odot \Phi_d(a_{\eta_{(2)}})\\
&= \sum a_{\eta_{(1)}}(c) \otimes a_{\eta_{(2)}}(d) \\
&= \Delta a_{\eta}(c,d) = (c\shuffle d, \eta).
\end{align*}

Hence, alternatively, the coproduct can be realized as the Hopf convolution product of the characters of the Hopf algebra $H$. The group inverse for any character $\Phi_{c}$ is defined as $\Phi_{c}^{\star -1} = \Phi_{c^{\shuffle^{-1}}} = \Phi_{c}\circ S$. It is not hard to see that the group of characters of the Hopf algebra $H$ and the shuffle group $M$ are isomorphic.
\begex Suppose $X=\{x_1,x_2\}$. Let $c = 1-x_1$ and $d = 1+x_1x_2$ $\in \allseries{}$. The shuffle product $c\shuffle d$ is computed directly from \rref{eqn:shuff_prod_def}  as $c\shuffle d = 1 + x_1x_2 - 2x_{1}^2x_2 - x_1x_2x_1$. The objective is to find only $(c\shuffle d,x_1x_2x_1) = \Delta a_{x_1x_2x_1}(c,d)$ using Algorithm~\ref{alg:cop}.

\noindent
(1) Recursively apply the map $\mu$ to compute the partition of the word $x_1x_2x_1$:
\begin{align*}
\mu(x_1x_2x_1) &= (x_1 \otimes \emptyset + \emptyset \otimes x_1)\mu(x_2x_1) \\
&= (x_1 \otimes \emptyset + \emptyset \otimes x_1)(x_2 \otimes \emptyset + \emptyset \otimes x_2)\mu(x_1) \\
&= (x_1 \otimes \emptyset + \emptyset \otimes x_1)(x_2 \otimes \emptyset + \emptyset \otimes x_2) (x_1 \otimes \emptyset + \emptyset \otimes x_1)\\
&= (x_1 \otimes \emptyset + \emptyset \otimes x_1)(x_2x_1 \otimes \emptyset + x_2 \otimes x_1 + x_1 \otimes x_2 + \emptyset \otimes x_2x_1) \\
&= x_1x_2x_1 \otimes \emptyset + x_1x_2 \otimes x_1 + x_1^2 \otimes x_2 + x_1 \otimes x_2x_1 + x_2x_1 \otimes x_1 + x_2 \otimes x_1^2 + \\
&\hspace*{0.2in} x_1 \otimes x_1x_2 + \emptyset \otimes x_1x_2x_1.
\end{align*}

\noindent
(2) Compute the coproduct:
\begin{align*}
\Delta a_{x_1x_2x_1}(c,d) &=  (c,x_1x_2x_1)(d,\emptyset) + (c,x_1x_2)(d,x_1) + (c,x_1^2)(d,x_2)+ (c,x_1)(d,x_2x_1) + \\
& \hspace*{0.4in}(c,x_2x_1)(d,x_1) + (c,x_2)(d,x_1^2) + (c,x_1)(d,x_1x_2) + (c,\emptyset)(d,x_1x_2x_1) \\
& = (0)(1) + (0)(0) + (0)(0) + (-1)(0) + (0)(0)+ (0)(0)+ (-1)(1) + (1)(0) \\
& = -1.
\end{align*}
Therefore, $(c \shuffle d, x_1x_2x_1)=-1$ as computed from the direct shuffle product calculation.
\endex 

A key observation is that Algorithm~\ref{alg:redcop} can be utilized to compute the Wiener-Fliess composition product \rref{eqn:mixcomp}. Specifically, if $\bar{c} \in \allproperseries{\ell}$, define the corresponding group element in $M$ as $c \triangleq \doubleone + \bar{c}$. If $\tilde{\eta} = \tilde{x}_{i_1}\tilde{x}_{i_2}\cdots \tilde{x}_{i_s} \in \allcommutingwords$ and $\zeta \in \allwords$, then 
\begin{align*}
(\bar{c}^{\shuffle \tilde{\eta}},\zeta) &= (\Delta^{\prime \circ (s-1)}a_{\zeta})(c_{i_1},c_{i_2},\ldots,c_{i_s}),
\end{align*}
where $(\Delta^{\prime \circ (s-1)}a_{\zeta})$ denotes the composition of the reduced coproduct map with itself $s-1$ times and then applied to the coordinate map $a_{\zeta}$. Computationally, this boils down to splitting the word $\zeta$ into all possible $s$ subwords, say $\zeta = \alpha_{1}\alpha_{2}\cdots\alpha_{s}$, where $\alpha_{i} \in X^{+}$, and then finding the Hadamard product of the coefficients corresponding to each subword with respect to the proper part of the series in the argument. That is,
\begin{align*}
(\bar{c}^{\shuffle \tilde{\eta}},\zeta) &= \sum_{\substack{\alpha_{1},\ldots,\alpha_{s} \in X^{+} \\ \zeta \in \alpha_{1}\shuffle\cdots\shuffle\alpha_{s}}}(\bar{c}_{i_1},\alpha_{1}) \odot (\bar{c}_{i_2},\alpha_{2}) \odot \cdots \odot (\bar{c}_{i_s},\alpha_{s}).
\end{align*}

The extension of this framework to the computation of the Wiener-Fliess composition product is described next. Let $\tilde{X} = \{\tilde{x}_1,\ldots,\tilde{x}_\ell\}$ be the commuting alphabet and $\tilde{X}^{\ast}$ the set of commuting words with $\tilde{X}^+ = \tilde{X}^{\ast}\setminus\{\emptyset\}$. Hence, $\;\forall \tilde{\eta} = \tilde{x}_{i_1}\tilde{x}_{i_2}\cdots\tilde{x}_{i_s} \in \allcommutingwords$ define the computational operators on the Hopf algebra $H$ as $\chi_{\tilde{\eta}} : H \longrightarrow H$ such that $a_{\eta} \mapsto \chi_{\tilde{\eta}}\left(a_{\eta}\right)$ (where $\eta \in \allwords$) and
\begin{align*}
\chi_{\tilde{\eta}}a_{\eta}(c) = \Delta^{\prime \circ (s-1)}a_{\eta}\left(c_{i_1},c_{i_2},\ldots,c_{i_s}\right) = \left(c^{\shuffle \tilde{\eta}},\eta\right) = \left(\tilde{\eta}\mixcomp \bar{c},\eta\right),
\end{align*}  
where $c \in M, \bar{c} \in \allproperseriesell$ and $c = \doubleone + \bar{c}$. If $d \in \allcommutingseriesX{k}$, then the Wiener-Fliess composition $d\mixcomp \bar{c}$ can be computed as  
\begin{align*}
(d \mixcomp \bar{c},\eta) &= \left(\left[\left(d,\emptyset\right)\epsilon + \sum_{\tilde{\eta} \in \tilde{X}^+}\left(d,\tilde{\eta}\right)\chi_{\tilde{\eta}}\right]a_{\eta}\right)(\doubleone + \bar{c}). 
\end{align*}
The framework for computing the Wiener-Fliess composition product for the non-proper case when $\bar{c} \in \allseries{\ell}\setminus\allproperseriesell $ and $d \in \allcommutingpolyX{k}$ requires more careful attention. Consider the case where $\bar{c}$ is non-proper such that $\left(\bar{c}_i,\emptyset\right) = r_i \neq 0 \; \forall i = 1,\ldots,\ell$. The corresponding group element of $\bar{c}$ in $M$ is $c$, where $c_i = \left(\frac{1}{r_i}\right)\bar{c}_i \; \forall i = 1,\ldots,\ell$. If $\tilde{\eta} = \tilde{x}_{i_1}\tilde{x}_{i_2}\cdots \tilde{x}_{i_s} \in \allcommutingwords$ and $\zeta \in \allwords$, then 
\begin{align*}
(\bar{c}^{\shuffle \tilde{\eta}},\zeta) &= \left(\prod_{j=1}^sr_{i_j}\right)\left(\Delta^{\circ (s-1)}a_{\zeta}\right)\left(c_{i_1},c_{i_2},\ldots,c_{i_s}\right),
\end{align*}
where $(\Delta^{\circ (s-1)}a_{\zeta})$ denotes the composition of the coproduct map with itself $s-1$ times and then applied to the coordinate map $a_{\zeta}$. Hence, $\;\forall \tilde{\eta} = \tilde{x}_{i_1}\tilde{x}_{i_2}\cdots\tilde{x}_{i_s} \in \allcommutingwords$, define the computational operator $\hat{\chi}_{\tilde{\eta}}$ on the Hopf algebra $H$ viz. $\hat{\chi}_{\tilde{\eta}} : H \longrightarrow H$ such that $a_{\eta} \mapsto \hat{\chi}_{\tilde{\eta}}\left(a_{\eta}\right)$ (where $\eta \in \allwords$) and
\begin{align*}
\hat{\chi}_{\tilde{\eta}}a_{\eta}(c) = \left(\prod_{j=1}^s\left(\bar{c}_{i_j},\emptyset\right)\right)\left(\Delta^{\circ (s-1)}a_{\zeta}\right)\left(c_{i_1},c_{i_2},\ldots,c_{i_s}\right) = \left(c^{\shuffle \tilde{\eta}},\eta\right) = \left(\tilde{\eta}\mixcomp \bar{c},\eta\right),
\end{align*}  
where $\bar{c} \in \allseries{}\setminus\allproperseriesell$ such that $\left(\bar{c}_i,\emptyset\right) \neq 0\; \forall i = 1,\cdots,\ell$, and $c \in M$ is the corresponding group element of $\bar{c}$. Therefore, let $\bar{c} \in \allseries{\ell}\setminus\allproperseriesell $ and $d \in \allcommutingpolyX{k}$. The Wiener-Fliess composition $d\mixcomp \bar{c}$ can be computed as
\begin{align*}  
(d \mixcomp \bar{c},\eta) &= \left(\left[\left(d,\emptyset\right)\epsilon + \sum_{\tilde{\eta} \in \supp\left(d\right)}\left(d,\tilde{\eta}\right)\hat{\chi}_{\tilde{\eta}}\right]a_{\eta}\right)(c).
\end{align*}
The case when $d \in \allcommutingpolyX{k}$ and $\bar{c} \in \allseries{\ell}\setminus\allproperseriesell$ is a non-proper series such that $\exists j \in \{1,\ldots,\ell\}: \;\left(c_j,\emptyset\right) = 0$ does not fit well into the framework based on the Hopf algebra $H$. Observe that this scenario can arise only if $\ell > 1$.

\section{Chen-Fliess Series under Static Output Feedback}\label{sec:static-feedback}
\begin{figure}[t]
	\begin{center}
		\includegraphics[width=0.5\textwidth]{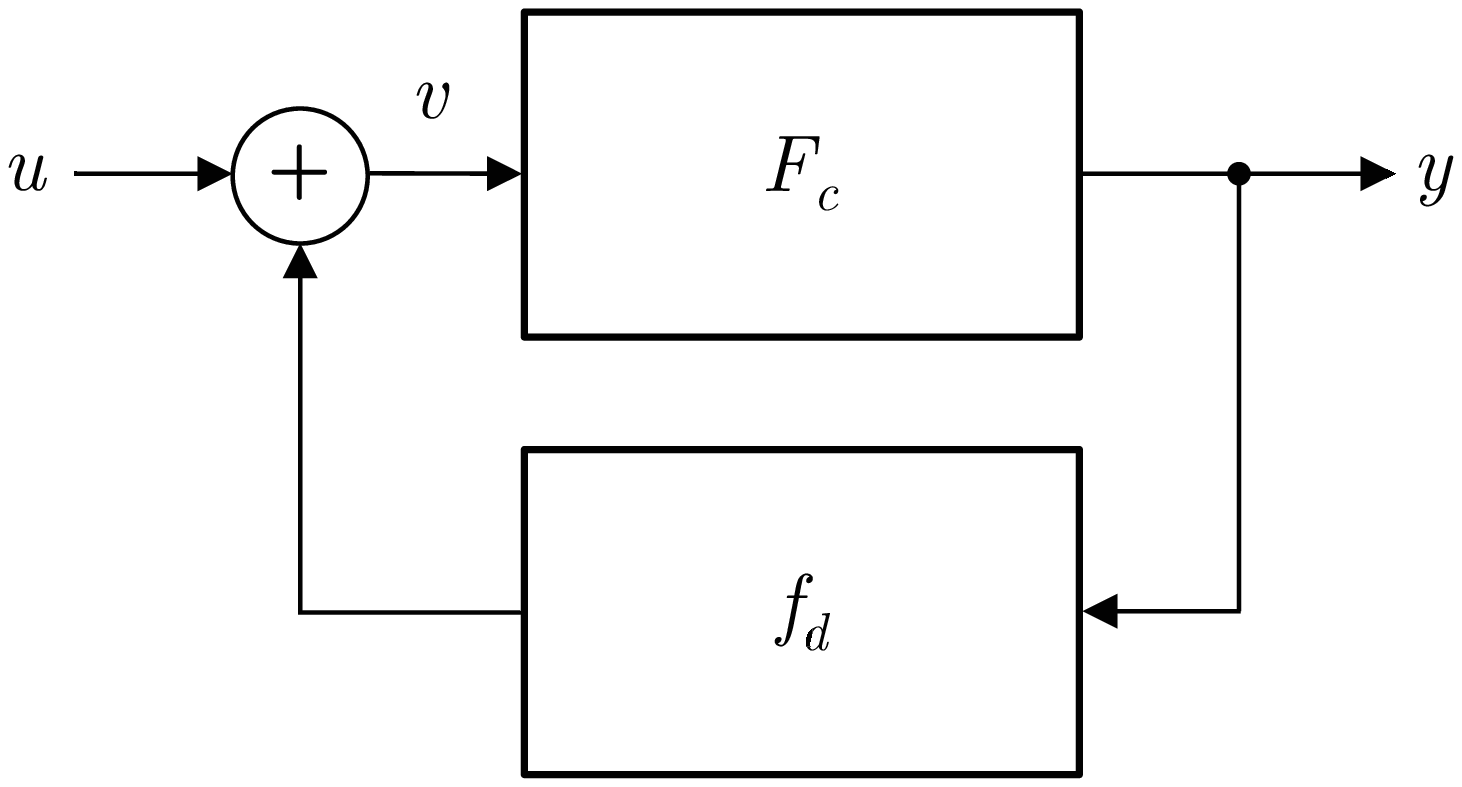}
	\end{center}
	\caption{Fliess operator $F_c$ with static output feedback $f_d$.}
	\label{fig:static_output_feedback}
\end{figure}

\paragraph*{}
Assume $\lvert X \rvert = m + 1$ and $\lvert \tilde{X} \rvert = \ell$. Let $F_c$ be a Chen-Fliess series with a generating series $c \in \allseries{\ell}$. Assume it is interconnected with a static formal map $f_d$ with generating series $d \in \allcommutingseriesX{m}$ in the additive output feedback configuration shown in Figure~\ref{fig:static_output_feedback} satisfying either of the following conditions:
\begin{enumerate}
	\item The series $c$ is proper. 
	\item $d$ is only a polynomial.
\end{enumerate}

The first objective of this section is to show that the closed-loop system always has a Chen-Fliess series representation, say $y=F_e[u]$, where $e\in\allseriesell$.
If this is the case, then necessarily
\begin{flalign*}
F_e[u] &= y \\
&= F_c[u + f_d(y)] \\
&= F_c[u + f_d \circ F_e[u]] \\
&= F_{c \modcomp (d \mixcomp e)_{\delta}}[u]
\end{flalign*}
for any admissible $u$. From the uniqueness of generating series, the series $e$ has to satisfy the fixed point equation
\begin{flalign}
e &= c \modcomp (d \mixcomp e)_{\delta}.
\label{eqn: fixed_point}
\end{flalign}
Observe that $e$ must be a proper series whenever $c$ is proper. It follows directly from the definition of the
mixed composition product that for all $w \in \allseriesXpri{k}$, the series $c\modcomp w_{\delta} \in \allseriesXpri{\ell}$ is proper if and only if $c \in \allproperseries{\ell}$, where $k = \lvert X \rvert -1$. The following lemmas will be used to show that \rref{eqn: fixed_point} always has a unique fixed point in both cases.

\begle\label{lem:Q_con} If $c \in \allproperseries{\ell}$ and $d \in \allcommutingseriesX{m}$, then the map
$Q_{c,d}: \allproperseries{\ell} \longrightarrow \allproperseries{\ell}:e \mapsto c \modcomp (d\mixcomp e)_{\delta}$ is a
strong contraction map in the ultrametric topology on the space $\allproperseries{\ell}$.
\endle

\begpr
First observe that $\kappa(h_{\delta}) = \kappa (h)$, $\forall h \in \allseries{\ell}$.
Now define two maps, $d_{\mixcomp,\small{\delta}} : e \mapsto (d \mixcomp e)_{\delta}$ and
$c_{\modcomp} : \: f \mapsto c \modcomp f_{\delta}$, where $f \in \allseries{m} $.
Note that $Q_{c,d}(e) = (c_{\modcomp} \circ d_{\mixcomp, \small{\delta}}) (e) $.
It is known that $c_{\modcomp}$ is a strong contraction map in the ultrametric topology \cite{Gray-Li_05}, so it only needs to be shown that $d_{\mixcomp,\small{\delta}}$ is at least a non-expansive map.

Consider first the case where $\overline{\omega}(d) = 1$. By Theorem~\ref{th:Wiener-Fliess-contraction},
$
\kappa(d_{\mixcomp,\small{\delta}}(e)) \leq \kappa(e).
$
Therefore, $d_{\mixcomp,\small{\delta}}$ is a weak contraction map.

Consider next the case where $\overline{\omega}(d) > 1$. Since $e \in \allproperseries{\ell}$ then $\orde(e) \geq 1$.
Therefore, $\kappa(e) \leq \sigma$ with $\sigma \in ]0,1[$. By Theorem~\ref{th:Wiener-Fliess-contraction},
$
\kappa(d_{\mixcomp,\small{\delta}}(e)) \leq \sigma\kappa(e).
$
Hence, $d_{\mixcomp,\small{\delta}}$ is a strong contraction map.
\endpr 

The counterpart of Lemma~\ref{lem:Q_con} for the case where $d$ is a polynomial but $c$ is allowed to be an arbitrary formal series not necessarily proper is proven next.

\begle\label{lem:Q_con_2} If $c \in \allseries{\ell}$ and $d \in \allcommutingpolyX{m}$, then the map
$\tilde{Q}_{c,d}: \allseries{\ell} \longrightarrow \allseries{\ell}:e \mapsto c \modcomp (d\mixcomp e)_{\delta}$ is a strong contraction map in the ultrametric topology on the space $\allseries{\ell}$.
\endle

\begpr 
Define the maps, $\tilde{d}_{\mixcomp,\small{\delta}} : e \mapsto (d \mixcomp e)_{\delta}$ and
$c_{\modcomp} : \: f \mapsto c \modcomp f_{\delta}$, where $f \in \allseries{m} $.
Note that $\tilde{Q}_{c,d}(e) = (c_{\modcomp} \circ d_{\mixcomp, \small{\delta}}) (e) $.
Since $\kappa(h_{\delta}) = \kappa (h)$, $\forall h \in \allseries{\ell}$, from Theorem~\ref{thm:WF_contraction}, $d_{\mixcomp,\small{\delta}}$ is a weak contraction map in ultrametric topology. As seen in Lemma~\ref{lem:Q_con}, the map $c_{\modcomp}$ is a strong contraction map in ultrametric space. Therefore, $\tilde{Q}_{c,d}(e) = (c_{\modcomp} \circ d_{\mixcomp, \small{\delta}}) (e) $ is a strong contraction map in the ultrametric topology on the space  $\allseries{\ell}$.
\endpr 

The following fixed point theorem establishes the first main result of the section, which follows subsequently.

\begth\label{thm:fixed_point} Let $X$ be a noncommutative alphabet and $\tilde{X}$ a commutative alphabet 
such that $\abs{X} = m+1$ and $\abs{\tilde{X}} = \ell$. The following statements are true:
\begin{enumerate}
	\item Given a proper series  $c \in \allproperseriesell$ and $d \in \allcommutingseriesX{m}$, the series $c \modcomp (-d \mixcomp c)_{\delta}^{-1}\in \allproperseriesell$ is a unique fixed point of the map $Q_{c,d}$ defined in Lemma~\ref{lem:Q_con}.
	\item Given a non-proper series $c \in \allseries{\ell}\setminus\allproperseriesell$ and $d \in \allcommutingpolyX{m}$, the series $c \modcomp (-d \mixcomp c)_{\delta}^{-1}\in \allseries{}\setminus\allproperseriesell$ is a unique fixed point of the map $\tilde{Q}_{c,d}$ defined in Lemma~\ref{lem:Q_con_2}. 
\end{enumerate}
\endth

\begpr Let $c \in \allproperseriesell$ and $d \in \allcommutingseriesX{m}$. If $e:= c \modcomp (-d \mixcomp c)^{-1}_{\delta}$, then
\begin{align*}
Q_{c,d}(e) &= c \modcomp (d \mixcomp e)_{\delta} \\
&= c \modcomp (d \mixcomp (c \modcomp (-d \mixcomp c)_{\delta}^{-1}))_{\delta}.
\end{align*}
Applying Theorem~\ref{thm:mix-assoc} yields
\begin{align*}
Q_{c,d}(e) &= c \modcomp ((d \mixcomp c) \modcomp (-d \mixcomp c)_{\delta}^{-1})_{\delta} \\
&= c \modcomp (-d \mixcomp c)^{-1}_{\delta} = e.
\end{align*}
Therefore, $c \modcomp (-d \mixcomp c)^{-1}_{\delta}$ is the unique fixed point of $Q_{c,d}$. Note that the uniqueness is guaranteed as all ultrametric spaces are Hausdorff spaces. The proof for the case when $c \in \allseries{\ell}\setminus\allproperseriesell$ and $d \in \allcommutingpolyX{m}$ is similar.
\endpr

\begth \label{thm:static_feedb}
Let $\abs{X} = m+1$ and $\abs{\tilde{X}}=\ell$. Assume either of the following holds:
\begin{enumerate}
	\item A proper series $c \in \allproperseries{\ell}$ and $d \in \allcommutingseriesX{m}$.
	\item A non-proper series $c \in \allseries{\ell}\setminus\allproperseriesell$ and $d \in \allcommutingpolyX{m}$.
\end{enumerate} 
Then the generating series for the closed-loop system in Figure~\ref{fig:static_output_feedback} is
the Wiener-Fliess feedback product $c\hat{@}d:= c \modcomp (-d \mixcomp c)^{-1}_{\delta}$.
\endth

The computation of $-d\mixcomp c$ can be performed via the coproduct of the Hopf algebra of
the shuffle group as described in Section~\ref{sec:HA-shuffle-group}. The group inverse $(-d \mixcomp c)^{-1}_{\delta}$ can be computed via the antipode of the Fa\'{a} di Bruno type Hopf algebra corresponding to the group $(\allseriesdelta{\ell},\circ,\delta)$. (A particularly efficient algorithm appears in \cite{Ebrahimi-Fard-Gray_17}.) Hence, the calculation of the generating series for the static feedback case is an interplay between these two very distinct Hopf algebras.

The notion that feedback can be described mathematically as a transformation group acting on the plant is well established in control theory~\cite{Brockett_78}. The following theorem describes the situation in the present context.
\begth
The Wiener-Fliess feedback product is a 
\begin{enumerate}
	\item right group action by the additive group $\left(\allcommutingseriesX{m},+,0\right)$
	on the set $\allproperseriesell$;
	\item right group action by the additive group $\left(\allcommutingpolyX{m},+,0\right)$
	on the set $\allseries{\ell}\setminus\allproperseriesell$,
\end{enumerate}
where $\lvert X \rvert = m+1$ and $\lvert \tilde{X} \rvert = \ell$.
\endth

\begpr
Let $d_{1},d_{2} \in \allcommutingseriesX{m}$ and $c \in \allproperseriesell$. It needs to be proved that
\begin{align*}
(c\hat{@}d_{1})\hat{@}d_{2} &= c\hat{@}(d_{1} + d_{2}).
\end{align*}
From Theorem~\ref{thm:static_feedb} observe that
\begin{align*}
(c\hat{@}d_{1})\hat{@} d_{2} &= (c\hat{@}d_{1}) \modcomp (-d_{2} \mixcomp (c\hat{@}d_{1}))_{\delta}^{-1} \\
&= (c \modcomp (-d_{1} \mixcomp c)_{\delta}^{-1}) \modcomp (-d_{2} \mixcomp (c\hat{@}d_{1}))_{\delta}^{-1}.
\end{align*}
Applying Theorem~\ref{thm:modcomp_grp} and then Theorem~\ref{thm:static_feedb} gives
\begin{align*}
(c\hat{@}d_{1})\hat{@}d_{2}
&= c \modcomp \big[(-d_{1}\mixcomp c)_{\delta}^{-1} \circ (-d_{2} \mixcomp (c\hat{@}d_{1}))_{\delta}^{-1}\big] \\
&= c \modcomp \big[(-d_{2} \mixcomp (c\hat{@}d_{1}))_{\delta} \circ (-d_{1}\mixcomp c)_{\delta}\big]^{-1} \\
&= c \modcomp\big[(-d_{2} \mixcomp (c \modcomp (-d_{1} \mixcomp c)_{\delta}^{-1}))_{\delta} \circ (-d_{1}\mixcomp c)_{\delta}\big]^{-1}.
\end{align*}
In light of Theorem~\ref{thm:mix-assoc},
\begin{align*}
(c\hat{@}d_{1})\hat{@}d_{2}
&= c \modcomp \bigg[\big((-d_{2}\mixcomp c)\modcomp (-d_{1} \mixcomp c)_{\delta}^{-1}\big)_{\delta} \circ \big(-d_{1}\mixcomp c \big)_{\delta}\bigg]^{-1}.
\end{align*}
Expanding the group product of $(\allseriesdelta{m},\circ)$, it follows that
\begin{align*}
(c\hat{@}d_{1})\hat{@}d_{2}
&= c \modcomp \bigg[\bigg(\bigg(\big((-d_{2}\mixcomp c)\modcomp (-d_{1} \mixcomp c)_{\delta}^{-1}\big) \modcomp \big(-d_{1}\mixcomp c \big)_{\delta}\bigg) + \big(-d_{1}\mixcomp c \big)\bigg)_{\delta} \bigg]^{-1}.
\end{align*}
Finally, from Theorem~\ref{thm:modcomp_grp},
\begin{align*}
(c\hat{@}d_{1})\hat{@}d_{2} &= c \modcomp [-d_{1}\mixcomp c + (-d_{2} \mixcomp c)]_{\delta}^{-1},
\end{align*}
so that via the left linearity of Wiener-Fliess composition,
\begin{align*}
(c\hat{@}d_{1})\hat{@}d_{2} &= c \modcomp [-(d_{1}+d_{2}) \mixcomp c]_{\delta}^{-1} = c\hat{@}(d_{1}+d_{2}).
\end{align*}
The proof is analogous for the case when $c$ is non-proper and $d_1,d_2 \in \allcommutingpolyX{m}$.
\endpr 

It is worth noting that for dynamic output feedback the transformation group is $\left(\allseries{m},+\right)$, while here it is  $(\allcommutingseriesX{m},+)$ (or $(\allcommutingpolyX{m},+)$) that plays this role. The final theorem states that a single-input single-output (SISO) nonlinear input-output system with relative degree has its relative degree left invariant under static output feedback. This fact is well known in the state space setting~\cite{Isidori_95}.
\begth\label{thm:static_fdb_rel_deg}
Let $X = \{x_0,x_1\}$ and $c \in \allseries{}$ have relative degree. If either of the following conditions hold:
\begin{enumerate}
	\item $c$ is proper and $d \in \allcommutingseriessingvar$;
	\item $c$ is non-proper and $d \in \allcommutingpolysingvar$,
\end{enumerate}
then $c \hat{@} d$ has relative degree equal to that of $c$.
\endth

\begpr The proof follows from the formula in Theorem~\ref{thm:static_feedb} and the relative
degree properties summarized in Table~$2$ of \cite{Gray_GS_relative degree}.
\endpr 

Observe that in the SISO case, the Wiener-Fliess composition product of a non-proper Chen-Fliess series $c$ with relative $r_c$ and a commutative polynomial $d \in \allcommutingpolyX{}$ can fail to have a well-defined relative degree as demonstrated in Example~\ref{ex:WF_rel_deg_fail}. However, the static feedback configuration of the non-proper series $c$ with the commutative polynomial $d$ always has well defined relative degree $r_c$ as proven in Theorem~\ref{thm:static_fdb_rel_deg}.     
\begex
Consider a normalized forced pendulum equation
\begeq \label{eqn:pend}
\ddot{\theta} + \sin \theta = u
\endeq
with input $u$, angular displacement $\theta$, and output $y=\theta$.
Under the feedback law $u = v + \sin \theta$, the system is transformed into a double integrator
$\ddot{\theta} = v$. For example, with $\theta(0)=0$ and $\dot{\theta}(0)=1$, the closed-loop system
is described by
\begdi
y(t)=t+\int_{0}^t\int_0^{\tau_2} v(\tau_1)\,d\tau_1\,d\tau_2,
\enddi
or equivalently, $y=F_{c\hat{@}d}[v]$ with $c\hat{@}d=x_0+x_0x_1$. Clearly, the
series has relative degree two. The same result can be established via Theorem~\ref{thm:static_feedb}.
The following computations were all done via Mathematica. It is easily checked that the open-loop system $y=F_c[u]$ has the generating series
\begin{align*}
c &= x_0 + x_0x_1 - x_0^3 - x_0^3x_1 + 2 x_0^5 + 4 x_0^5 x_1 + 2 x_0^4x_1x_0 + x_0^3x_1x_0^2 + \cdots
\end{align*}
and has relative degree 2 as expected.
The sinusoidal static output feedback map has generating series $d \in \allcommutingseriessingvar{}$ given by
\begin{align*}
d &= \tilde{x}_{1} - \frac{1}{3!}\tilde{x}_1^3 + \frac{1}{5!}\tilde{x}_1^5 - \frac{1}{7!}\tilde{x}_1^7 +  \cdots
\end{align*}
Using the computational methods described above and computing
the composition antipode for words up to length four, it is found that
\begin{flalign*}
c\hat{@}d &\approx  x_0 + x_0x_1 + \mathcal{O}(x_0^6).
\end{flalign*}
The terms $\mathcal{O}(x_{0}^6)$ are the error terms due to the need to truncate all the underlying series at each step of the calculation in the Wiener-Fliess feedback product formula. The order of these error terms can be increased but at a significant computational cost.
\endex

\section{Local Convergence of Wiener-Fliess Composition Product}\label{sec:loc_conv}
\paragraph*{}
The goal of this section is to prove that local convergence is preserved under Wiener-Fliess composition product, for which the following lemma is essential.

\begle\label{lemma:loc}\cite{GS_thesis} Let $e \in \allseriesLC{}$ be a proper series such that $\lvert(e,\zeta)\rvert \leq K_e M_e ^{\lvert \zeta \rvert-j} \lvert\zeta\rvert! \; \forall \zeta \in X^+$, where $j \geq 0$. Then $e^{\shuffle n}$ is a proper and locally convergent series $\forall n \geq 1$ such that $\lvert(e^{\shuffle n},\zeta)\rvert \leq K_e^n M_e ^{\lvert\zeta\rvert-nj} \lvert\zeta\rvert! \binom{\lvert\zeta\rvert-1}{n-1} \; \forall \zeta \in X^+$.
\endle

The following theorem addresses the preservation of local convergence under the Wiener-Fliess composition product. Only the case where $c$ is a noncommutative formal proper series, as defined in Theorem~\ref{thm:Wiener-Fliess-product}, is considered here. The case where $d$ is a commutative polynomial is addressed in Section~\ref{sec:glo_conv}. The theorem in spirit has appeared in \cite{Gray_Thitsa_IJC12}, where it was proved using exponential generating functions. In this work, an alternate proof is provided using only elementary combinatorics.   

\begth\label{thm:loc_wiener} Let $d \in \allcommutingseriesLCX{}$ with $\abs{\tilde{X}} = k$ and $\abs{(d,\eta)} \leq K_dM_d^{\abs{\eta}} \; \forall \eta \in \tilde{X}^{\ast}$. If $c \in \allseriespLC{k}$ such that $\abs{(c_i,\zeta)} \leq K_{c_i}M_{c_i}^{\abs{\zeta}}\abs{\zeta}! \; \forall \zeta \in \allwords$, then $d\mixcomp c \in \allseriesLC{}$ with
\begin{align*}
\abs{(d\mixcomp c, \eta)} &\leq 
\begin{cases}
\left(\frac{kK_dK_cM_d}{1+kK_cM_d}\right)\left[M_c\left(1+kK_cM_d\right)\right]^{\abs{\eta}}\abs{\eta}! &\quad \text{if } \eta \neq \emptyset \\
K_d &\quad \text{if } \eta = \emptyset, 
\end{cases}
\end{align*}
where $K_c = \max\limits_{i = 1,\ldots,k}K_{c_i}$ and $M_c = \max\limits_{i = 1,\ldots,k}M_{c_i}$.
\endth

\begpr Observe
\begin{align*}d \mixcomp c &= \sum_{\eta \in \tilde{X}^{\ast}} (d,\eta) c^{\shuffle \eta} \\
&\leq \sum_{\eta \in \tilde{X}^{\ast}} K_dM_d^{\abs{\eta}} c^{\shuffle \eta} \\
&= \sum_{n=0}^{\infty} K_dM_d^n \sum_{\eta \in \tilde{X}^n} c^{\shuffle \eta} \\
&= \sum_{n=0}^{\infty} K_dM_d^n \sum_{\substack{i_1,\ldots,i_k \geq 0 \\i_1 + \cdots + i_k = n}} \left(\tilde{x}_1^{i_1}\cdots\tilde{x}_k^{i_k}\right) \mixcomp c \\
&= \sum_{n=0}^{\infty} K_dM_d^n \sum_{\substack{i_1,\ldots,i_k \geq 0 \\i_1 + \cdots + i_k = n}} c_1^{\shuffle i_1}\shuffle c_2^{\shuffle i_2} \shuffle \cdots \shuffle c_k^{\shuffle i_k}. 
\end{align*}
Since $\binom{n}{i_1\cdots i_k} \geq 1$,
\begin{align*}
d \mixcomp c &\leq \sum_{n=0}^{\infty} K_dM_d^n \sum_{\substack{i_1,\ldots,i_k \geq 0 \\i_1 + \cdots + i_k = n}} \binom{n}{i_1\cdots i_k} c_1^{\shuffle i_1}\shuffle c_2^{\shuffle i_2} \shuffle \cdots \shuffle c_k^{\shuffle i_k} \\
&= \sum_{n=0}^{\infty} K_dM_d^n \left(\sum_{j=1}^k c_j \right)^{\shuffle n} \\
&= K_d \sum_{n=0}^{\infty} \left(M_d\sum_{j=1}^k c_j \right)^{\shuffle n}.
\end{align*}
Note that $M_d\sum_{j=1}^kc_j$ is a proper series. Hence, $\abs{(d\mixcomp c, \emptyset)} \leq K_d$. Now let $\eta \in  X^{\alpha}$, where $\alpha \in \nat$. Then,
\begin{align*}
(d\mixcomp c, \eta) &\leq K_d \left(\left[\sum_{n=1}^{\alpha} \left(M_d\sum_{j=1}^k c_j \right)\right]^{\shuffle n},\eta \right) \\
&= K_d \sum_{n=1}^{\alpha}M_d^n \left(\left(\sum_{j=1}^k c_j\right)^{\shuffle n},\eta \right). 
\end{align*}
Observe that $\left(\sum_{j=1}^k c_j,\zeta\right) \leq k K_cM_c^{\abs{\zeta}}\abs{\zeta}!$, where $K_c = \max\limits_{i = 1,\ldots,k}K_{c_i}$ and $M_c = \max\limits_{i = 1,\ldots,k}M_{c_i}$. 
Using Lemma \ref{lemma:loc} and the triangle inequality,
\begin{align*}
\abs{(d \mixcomp c, \eta)} &\leq K_d \left[ \sum_{n=1}^{\alpha} k^nK_c^nM_d^nM_c^{\alpha} \alpha! \binom{\alpha-1}{n-1}\right] \\
&= K_d M_c^{\alpha}\alpha!\left[\sum_{n=1}^{\alpha} k^nK_c^nM_d^n\binom{\alpha-1}{n-1}\right] \\
&= K_d \left(kK_cM_d\right) M_c^{\alpha} \left[\sum_{n=0}^{\alpha-1} \left(kK_cM_d\right)^n \binom{\alpha-1}{n}\right] \alpha! \\
&= K_d \left(kK_cM_d\right) M_c^{\alpha} \left(1 + kK_cM_d\right)^{\alpha-1} \alpha! \\
&= \left(\frac{kK_dK_cM_d}{1+kK_cM_d}\right)\left[M_c\left(1+kK_cM_d\right)\right]^{\alpha}\alpha! \\
&= \left(\frac{kK_dK_cM_d}{1+kK_cM_d}\right)\left[M_c\left(1+kK_cM_d\right)\right]^{\abs{\eta}}\abs{\eta}!
\end{align*}
as claimed, and therefore, $d \mixcomp c \in \allseriesLC{}$.
\endpr

\section{Global Convergence of Wiener-Fliess Composition Product}\label{sec:glo_conv}
\paragraph*{}
This section addresses the preservation of global convergence under the Wiener-Fliess composition product. That is, the Wiener-Fliess composition of a globally convergent commutative series $d$ and a noncommutative series $c$ in the Fr\'echet space $\serSinf{}$, $d\mixcomp c$, lies in the Fr\'echet space. Recall from Theorem~\ref{thm:Wiener-Fliess-product}, the definition is two-fold. This section considers both these cases in detail. However, the proofs of these convergence theorems need a few preliminary results. In particular, the proofs of global convergence involve the use of fractional powers of multinomial coefficients. Recall that the gamma function, $\Gamma\left(\cdot\right)$,  restricted to $\re_+$ is the analytic continuation of the factorial map on the non-negative integers~\cite{Abramowitz_Stegun}. Hence, the analytic continuation of the multinomial coefficient is defined in the following way.  

\begde If $\alpha \in \re_{\geq 0}$ and $i_i,i_2,\ldots,i_s \in \re_{\geq 0}$ such that $\sum_{j = 1}^s i_j = \alpha$, then
\begin{align*}
\binom{\alpha}{i_1\; i_2\cdots i_s} &= \frac{\Gamma\left(\alpha+1\right)}{\Gamma\left(i_1+1\right)\;\Gamma\left(i_2+1\right)\cdots\Gamma\left(i_s+1\right)}.
\end{align*}
\endde

The following lemma is central to proving that $\serSinf{m}$ is closed under the shuffle product.  
\begle\label{lem:multinom} Let $\alpha \in \re_{\geq 0}$ and $i_1,i_2,\ldots,i_v \in \re_{\geq 0}$ such that $\sum_{j=1}^v i_j = \alpha$. If $r \in \left]0,1\right]$, then
\begin{align*}
\binom{\alpha}{i_1\;i_2\cdots i_v}^r &\leq \left(\frac{K_{r}^v}{\tilde{K}_r}\right)\left(\frac{M_r}{2}\right)^\alpha \binom{r\alpha}{ri_1\;ri_2\cdots ri_v},
\end{align*}
where $K_r = \left(\left(\frac{2\pi}{\exp(2)}\right)^{1-r}r\right)^{\frac{1}{2}}$, $\tilde{K}_r = 2\left(\left(\frac{2\pi}{\exp(2)}\right)^{1-r}4\right)^{\frac{1}{2}}$, and $M_r = r^r$.
\endle

\begpr Observe
\begin{align*}
\binom{\alpha}{i_1\;i_2\cdots i_v}^r &= \frac{\left(\Gamma\left(\alpha+1\right)\right)^r}{\left(\Gamma\left(i_1+1\right)\right)^r\;\left(\Gamma\left(i_2+1\right)\right)^r\cdots\left(\Gamma\left(i_v+1\right)\right)^r}.
\end{align*}
Using the Lemma~\ref{lem:factorial},
\begin{align*}
\binom{\alpha}{i_1\;i_2\cdots i_v}^r &\leq \frac{2^{-r}\tilde{K}_r^{-1}\Gamma\left(r\alpha+1\right)}{K_r^{-1}M_r^{-i_1}\Gamma\left(ri_1+1\right)K_r^{-1}M_r^{-i_2}\Gamma\left(ri_2+1\right)\cdots K_r^{-1}M_r^{-i_v}\Gamma\left(ri_v+1\right)} \\
&= \left(\frac{K_{r}^v}{\tilde{K}_r}\right)\left(\frac{M_r}{2}\right)^\alpha \binom{r\alpha}{ri_1\;ri_2\cdots ri_v}.
\end{align*}
\endpr

The following theorem is known as the \emph{Neoclassical Inequality} and is an extension of the multinomial theorem extended to arbitrary positive fractional powers of non-negative reals. 
\begth\label{thm:neoclass}\cite{Lyons-Qian_02}  Let $r \in \left]0,1\right]$ and $m \in \mathbb{N}$. If $n \in \mathbb{N}_{0}$ and $x_1,x_2,\ldots,x_m \geq 0$, then
\begin{align*}
\sum_{\substack{i_1,\ldots,i_m \in \mathbb{N}_0 \\ i_1+i\cdots+i_m = n}}\binom{rn}{ri_1\cdots ri_m} x_1^{ri_1}\cdots x_m^{ri_m} &\leq \left(\frac{1}{r}\right)^{2\left(m-1\right)} \left(x_1+\cdots+x_m\right)^n.
\end{align*}
\endth 

If $r = 1$ above, then the inequality becomes an equality and reduces to the well known multinomial theorem albeit restricted to positive reals. The following theorem addresses the particular case where the Wiener-Fliess composition yields a series in $\allseriesGC{}$.

\begth\label{thm:WFcomp_GC} If $d \in \allcommutingpolyX{}$ with $\abs{\tilde{X}} = m$ and $c \in \allseriesGC{m}$, then $d\mixcomp c \in \allseriesGC{}$.
\endth 

\begpr Assume $d$ is a polynomial of degree $N \in \nat_0$. Since $d \in \allcommutingpolyX{}$, there exist constants $K_d,M_d >0$ such that $\abs{\left(d,\eta\right)} \leq \dfrac{K_dM_d^{\abs{\eta}}}{\abs{\eta}!} \; \forall \eta \in \bigcup_{i =0}^N \tilde{X}^i$. 
Observe
\begin{align*}
d \mixcomp c &= \sum_{\eta \in \supp(d)} (d,\eta) c^{\shuffle \eta} \\
&\leq \sum_{\eta \in \supp(d)}\frac{K_dM_d^{\abs{\eta}}} {\abs{\eta}!}c^{\shuffle \eta} \\
&= \sum_{n=0}^{N} \frac{K_dM_d^n}{n!} \sum_{\eta \in \tilde{X}^n} c^{\shuffle \eta} \\
&= \sum_{n=0}^{N} \frac{K_dM_d^n}{n!} \sum_{\substack{i_1,\ldots,i_k \geq 0 \\i_1 + \cdots + i_k = n}} \left(\tilde{x}_1^{i_1}\cdots\tilde{x}_k^{i_k}\right) \mixcomp c \\
&= \sum_{n=0}^{N} \frac{K_dM_d^n}{n!} \sum_{\substack{i_1,\ldots,i_k \geq 0 \\i_1 + \cdots + i_k = n}}  c_1^{\shuffle i_1}\shuffle c_2^{\shuffle i_2} \shuffle \cdots \shuffle c_k^{\shuffle i_k}. 
\end{align*}
Since $\binom{n}{i_1\cdots i_k} \geq 1$,
\begin{align*}
d \mixcomp c &\leq \sum_{n=0}^{N} \frac{K_dM_d^n}{n!} \sum_{\substack{i_1,\ldots,i_k \geq 0 \\i_1 + \cdots + i_k = n}} \binom{n}{i_1\cdots i_k} c_1^{\shuffle i_1}\shuffle c_2^{\shuffle i_2} \shuffle \cdots \shuffle c_k^{\shuffle i_k} \\
&= \sum_{n=0}^{N} \frac{K_dM_d^n \left(\sum_{j=1}^m c_j \right)^{\shuffle n}}{n!} \\
&= K_d \sum_{n=0}^{N} \frac{\left[M_d\left(\sum_{j=1}^m c_j \right)\right]^{\shuffle n}}{n!} \\
&\leq K_d \sum_{n=0}^{N} \left[M_d\left(\sum_{j=1}^m c_j \right)\right]^{\shuffle n}.
\end{align*}
If $c \in \allseriesGC{m}$, then $c_j \in \allseriesGC{} \; \forall j = 1,\ldots,m$. Hence, $ \sum_{j=1}^m c_i \in \allseriesGC{}$ by Theorem~\ref{thm:addition_conv}. Using Corollary~\ref{cor:shuffle_GC_pow}, $\left[M_d\left(\sum_{j=1}^m c_j \right)\right]^{\shuffle n} \in \allseriesGC{} \; \forall n\leq N$. Hence, by virtue of Theorem~\ref{thm:addition_conv},
\begin{align*}
\sum_{n=0}^{N} \left[M_d\left(\sum_{j=1}^m c_j \right)\right]^{\shuffle n} \in \allseriesGC{}.
\end{align*}
Therefore, 
\begin{align*}
d \mixcomp c &\leq K_d \sum_{n=0}^{N} \left[M_d\left(\sum_{j=1}^m c_j \right)\right]^{\shuffle n} \in \allseriesGC{}.
\end{align*}
\endpr

The following theorem addresses the convergence for the Wiener-Fliess composition of a locally convergent noncommutative proper series with a globally convergent commutative series $d$ .  
\begth\label{thm:wiener_fliess_global} Let $d \in \allcommutingseriesGCX{}$ with $\abs{\tilde{X}} = m$, growth constants $K_d,M_d >0$, and Gevrey order $\left(-1+\bar{s}\right)$ with $\bar{s} \in [0,1[$ such that $\abs{\left(d,\tilde{\eta}\right)} \leq K_dM_d^{\abs{\tilde{\eta}}}\left(\abs{\tilde{\eta}}!\right)^{-1+\bar{s}} \; \forall \tilde{\eta} \in \tilde{X}^{\ast}$. If $c$ is a proper series such that $c \in \serSinfR{m}{R} \cap \allproperseries{m}$, then $d \mixcomp c \in \serSinfR{}{R^{\prime}} \; \forall R^{\prime} = \epsilon R$, where $\epsilon \in ]0,1[$.  
\endth

\begpr Observe
\begin{align*}
d \mixcomp c &= \sum_{\eta \in \tilde{X}^{\ast}} (d,\eta) c^{\shuffle \eta} \\
&\leq \sum_{\eta \in \tilde{X}^{\ast}} K_dM_d^{\abs{\eta}} \left(\abs{\eta}!\right)^{-1+\bar{s}}c^{\shuffle \eta} \\
&= \sum_{n=0}^{\infty} K_dM_d^n\left(n!\right)^{-1+\bar{s}} \sum_{\eta \in \tilde{X}^n} c^{\shuffle \eta} \\
&= \sum_{n=0}^{\infty} \frac{K_dM_d^n}{\left(n!\right)^{1-\bar{s}}} \sum_{\substack{i_1\cdots i_k \geq 0 \\i_1 + \cdots + i_k = n}} \left(\tilde{x}_1^{i_1}\cdots\tilde{x}_k^{i_k}\right) \mixcomp c \\
&= \sum_{n=0}^{\infty} \frac{K_dM_d^n}{\left(n!\right)^{1-\bar{s}}} \sum_{\substack{i_1\cdots i_k \geq 0 \\i_1 + \cdots + i_k = n}} c_1^{\shuffle i_1}\shuffle c_2^{\shuffle i_2} \shuffle \cdots \shuffle c_k^{\shuffle i_k}. 
\end{align*}
Since $\binom{n}{i_1\cdots i_k} \geq 1$,
\begin{align*}
d \mixcomp c &\leq \sum_{n=0}^{\infty} \frac{K_dM_d^n}{\left(n!\right)^{1-\bar{s}}} \sum_{\substack{i_1\cdots i_k \geq 0 \\i_1 + \cdots + i_k = n}} \binom{n}{i_1\cdots i_k} c_1^{\shuffle i_1}\shuffle c_2^{\shuffle i_2} \shuffle \cdots \shuffle c_k^{\shuffle i_k} \\
&= \sum_{n=0}^{\infty} \frac{K_dM_d^n \left(\sum_{j=1}^m c_j \right)^{\shuffle n}}{\left(n!\right)^{1-\bar{s}}} \\
&= K_d \sum_{n=0}^{\infty} \frac{\left[M_d\left(\sum_{j=1}^m c_j \right)\right]^{\shuffle n}}{\left(n!\right)^{1-\bar{s}}}.
\end{align*}
Lemma \ref{lem:factorial} implies that 
\begdi
\left(\Gamma(n+1)\right)^{(1-\bar{s})} = n!^{(1-\bar{s})} \geq \frac{1}{\tilde{K}_{1-\bar{s}}}2^{-n}\Gamma((1-\bar{s})n+1), 
\enddi
where $\tilde{K}_{1-\bar{s}} = 2\left(\left(\frac{2\pi}{\exp(2)}\right)^{\bar{s}}4\right)^{\frac{1}{2}}$. Hence,
\begin{align*}
d\mixcomp c  &\leq K_d\tilde{K}_{1-\bar{s}} \sum_{n=0}^{\infty} \frac{\left[2M_d\left(\sum_{j=1}^m c_j \right)\right]^{\shuffle n}}{\Gamma\left((1-\bar{s})n+1\right)}.
\end{align*}
Define $\tilde{K}_d \triangleq K_d\tilde{K}_{1-\bar{s}}$ and $ \tilde{M}_d \triangleq 2M_d$. Therefore,
\begin{align*}
d\mixcomp c \leq \tilde{K}_d\sum_{n=0}^{\infty} \frac{\left[\tilde{M}_d\left(\sum_{j=1}^m c_j \right)\right]^{\shuffle n}}{\Gamma\left((1-\bar{s})n+1\right)},
\end{align*}
and consequently, 
\begin{align*}
\norminf{R^{\prime}}{d\mixcomp c} &\leq \tilde{K}_d \norminf{R^{\prime}}{\sum_{n=0}^{\infty} \frac{\left[\tilde{M}_d\left(\sum_{j=1}^m c_j \right)\right]^{\shuffle n}}{\Gamma\left((1-\bar{s})n+1\right)}}.
\end{align*}
Applying the triangle inequality,
\begin{align*}
\norminf{R^{\prime}}{d \mixcomp c} &\leq \tilde{K}_d \sum_{n=0}^{\infty} \frac{{\tilde{M}_d}^n\norminf{R^{\prime}}{\left(\sum_{j=1}^m c_j \right)^{\shuffle n}}}{\Gamma\left((1-\bar{s})n+1\right)}.
\end{align*}
In addition, $\norminf{R}{\left(\sum_{j=1}^m c_j \right)} \leq  \sum_{j=1}^m \norminf{R}{c_j} \leq m\norminf{R}{c} < \infty$. Therefore, $\sum_{j=1}^m c_j \in \serSinfR{}{R}$. Hence, $\forall\;R^{\prime} = \epsilon R$, where $\epsilon \in ]0,1[$, by virtue of Corollary~\ref{cor:shuffle_power_conv},
\begin{align*}
\norminf{R^{\prime}}{d\mixcomp c} &\leq \tilde{K}_d \sum_{n=0}^{\infty} \frac{\left(\frac{\tilde{M}_d}{1-\epsilon}\right)^n\norminf{R}{\left(\sum_{j=1}^k c_j \right)}^n}{\Gamma\left((1-\bar{s})n+1\right)} \\
&\leq \tilde{K}_d \sum_{n=0}^{\infty} \frac{\left(\frac{\tilde{M}_d}{1-\epsilon}\right)^n \left(m\norminf{R}{c}\right)^n}{\Gamma\left((1-\bar{s})n+1\right)} \\
&= \tilde{K}_d \sum_{n=0}^{\infty} \frac{\left[\frac{\tilde{M}_d}{1-\epsilon}\left(m\norminf{R}{c}\right)\right]^n}{\Gamma\left((1-\bar{s})n+1\right)} \\
& = \tilde{K}_d\;\mathbb{E}_{\left(1-\bar{s}\right),1}\left(\frac{\tilde{M}_dm\norminf{R}{c}}{1-\epsilon}\right).
\end{align*}
Since $\bar{s} \in [0,1[$, the Mittag-Leffler function $\mathbb{E}_{\left(1-\bar{s}\right),1}\left(\cdot\right)$ is an entire function. Note that $c \in \serSinfR{m}{R}$ and $\epsilon \in ]0,1[$ implies that $\left(\dfrac{\tilde{M}_cm\norminf{R}{c}}{1-\epsilon}\right) < \infty$. Therefore,
\begin{align*} 
\norminf{R^{\prime}}{d\mixcomp c} \leq \tilde{K}_d\;\mathbb{E}_{\left(1-\bar{s}\right),1}\left(\frac{\tilde{M}_dm\norminf{R}{c}}{1-\epsilon}\right) < \infty.
\end{align*} 
Hence, $d \in \allcommutingseriesGCX{}$ and $c \in \serSinfR{m}{R}\cap\allproperseries{m}$ ensure that $d\mixcomp c\; \in \serSinfR{}{R^{\prime}} \; \forall R^{\prime} = \epsilon R$, where $\epsilon \in ]0,1[$.
\endpr

Theorem~\ref{thm:wiener_fliess_global} establishes that the convergence of the Wiener-Fliess composition product $d\mixcomp c$ is limited by the convergence of the proper noncommutative series $c$, when the commutative series $d$ is globally convergent. The following addresses the convergence of the Wiener-Fliess composition product when the series $c$ lies in the Fr\'echet space. 
\begth \label{thm:WF_glo_conv_ser} If $d \in \allcommutingseriesGCX{}$ with $\abs{\tilde{X}} = m$ and $c \in \serSinf{m} \cap \allproperseries{m}$, then $d \mixcomp c \in \serSinf{}$.
\endth

\begpr Recall \begin{align*}
c \in \serSinf{m}\cap\allproperseries{m} &\Longleftrightarrow c \in \serSinfR{m}{R}\cap\allproperseries{m} \; \forall R > 0.
\end{align*}
Fix $\epsilon \in ]0,1[$ and $\forall R^{\prime} >0$ define $R = \left(\dfrac{1}{\epsilon}\right) R^{\prime}$. Since $d \in \allcommutingseriesGCX{}$, using Theorem~\ref{thm:wiener_fliess_global} gives $d \mixcomp c \in \serSinfR{}{R^{\prime}}$. Therefore,
\begin{align*}
&d \mixcomp c \in \serSinfR{}{R^{\prime}} \; \forall R^{\prime} > 0 \Longleftrightarrow d\mixcomp c \in \serSinf{} .
\end{align*}
\endpr

Theorem~\ref{thm:WF_glo_conv_ser} proved that the Wiener-Fliess composition preserves global convergence when the noncommutative series in the composition is proper. The next step is to revisit the question for the case of the Wiener-Fliess composition when the commutative series is restricted to being a polynomial.
\begth\label{thm:WF_global_poly} If $d \in \allcommutingpolyX{}$ and $c \in \serSinfR{m}{R}$, then $d \mixcomp c \in \serSinfR{}{R^{\prime}} \; \forall R^{\prime} = \epsilon R$, where $\epsilon \in ]0,1[$.  
\endth

\begpr Assume $d$ is a polynomial of degree $N \in \nat_0$. Since $d \in \allcommutingpolyX{}$, there exist constants $K_d,M_d >0$ such that $\abs{\left(d,\eta\right)} \leq \dfrac{K_dM_d^{\abs{\eta}}}{\abs{\eta}!}, \;\forall \eta \in \cup_{i =0}^N \tilde{X}^i$. Observe 
\begin{align*}
d \mixcomp c &= \sum_{\eta \in \supp(d)} (d,\eta) c^{\shuffle \eta} \\
&\leq \sum_{\eta \in \supp(d)}\frac{K_dM_d^{\abs{\eta}}} {\abs{\eta}!}c^{\shuffle \eta} \\
&= \sum_{n=0}^{N} \frac{K_dM_d^n}{n!} \sum_{\eta \in \tilde{X}^n} c^{\shuffle \eta} \\
&= \sum_{n=0}^{N} \frac{K_dM_d^n}{n!} \sum_{\substack{i_1,\ldots,i_k \geq 0 \\i_1 + \cdots + i_k = n}} \left(\tilde{x}_1^{i_1}\cdots\tilde{x}_k^{i_k}\right) \mixcomp c \\
&= \sum_{n=0}^{N} \frac{K_dM_d^n}{n!} \sum_{\substack{i_1,\ldots,i_k \geq 0 \\i_1 + \cdots + i_k = n}}  c_1^{\shuffle i_1}\shuffle c_2^{\shuffle i_2} \shuffle \cdots \shuffle c_k^{\shuffle i_k}. 
\end{align*}
Since $\binom{n}{i_1\cdots i_k} \geq 1$,
\begin{align*}
d \mixcomp c &\leq \sum_{n=0}^{N} \frac{K_dM_d^n}{n!} \sum_{\substack{i_1,\ldots,i_k \geq 0 \\i_1 + \cdots + i_k = n}} \binom{n}{i_1\cdots i_k} c_1^{\shuffle i_1}\shuffle c_2^{\shuffle i_2} \shuffle \cdots \shuffle c_k^{\shuffle i_k} \\
&= \sum_{n=0}^{N} \frac{K_dM_d^n \left(\sum_{j=1}^m c_j \right)^{\shuffle n}}{n!} \\
&= K_d \sum_{n=0}^{N} \frac{\left[M_d\left(\sum_{j=1}^m c_j \right)\right]^{\shuffle n}}{n!}. 
\end{align*}
Therefore, 
\begin{align*}
\norminf{R^{\prime}}{d\mixcomp c} &\leq K_d \norminf{R^{\prime}}{\sum_{n=0}^{N} \frac{\left[M_d\left(\sum_{j=1}^m c_j \right)\right]^{\shuffle n}}{n!}}\\
&\leq K_d \sum_{n=0}^{N} \frac{{M_d}^n\norminf{R^{\prime}}{\left(\sum_{j=1}^m c_j \right)^{\shuffle n}}}{n!}.
\end{align*}
In addition, $\norminf{R}{\left(\sum_{j=1}^m c_j \right)}\leq  \sum_{j=1}^m \norminf{R}{c_j} \leq m\norminf{R}{c} < \infty$. Therefore, $\sum_{j=1}^m c_j \in \serSinfR{}{R}$. Hence, $\forall\;R^{\prime} = \epsilon R$, where $\epsilon \in ]0,1[$, by virtue of Corollary~\ref{cor:shuffle_power_conv},
\begin{align*}
\norminf{R^{\prime}}{d\mixcomp c} &\leq K_d \sum_{n=0}^{N} \frac{\left(\frac{M_d}{1-\epsilon}\right)^n\norminf{R}{\left(\sum_{j=1}^k c_j \right)}^n}{n!} \\
&\leq K_d \sum_{n=0}^{N} \frac{\left(\frac{M_d}{1-\epsilon}\right)^n \left(m\norminf{R}{c}\right)^n}{n!} \\
&= K_d \sum_{n=0}^{N} \frac{\left[\frac{M_d}{1-\epsilon}\left(m\norminf{R}{c}\right)\right]^n}{n!} \\
&\leq K_d \sum_{n=0}^{\infty} \frac{\left(\frac{M_d}{1-\epsilon}\right)^n\norminf{R}{\left(\sum_{j=1}^k c_j \right)}^n}{n!} \\
&= K_d \exp\left(\frac{M_dm\norminf{R}{c}}{1-\epsilon}\right).
\end{align*}
Observe that $c \in \serSinfR{m}{R}$ and $\epsilon \in ]0,1[$ imply that $\left(\dfrac{\tilde{M}_cm\norminf{R}{c}}{1-\epsilon}\right) < \infty$. Therefore,
\begin{align*} 
\norminf{R^{\prime}}{d\mixcomp c} \leq  K_d \exp\left(\frac{M_dm\norminf{R}{c}}{1-\epsilon}\right) <  \infty.
\end{align*} 
Hence, $d \in \allcommutingpolyX{}$ and $c \in \serSinfR{m}{R}$ must yield $d\mixcomp c\; \in \serSinfR{}{R^{\prime}} \; \forall R^{\prime} = \epsilon R$, where $\epsilon \in ]0,1[$.
\endpr

Theorem~\ref{thm:WF_global_poly} proves that the convergence of the Wiener-Fliess composition product $d\mixcomp c$ is limited by the convergence of the noncommutative series $c$, when $d$ is restricted to being a polynomial. The result is analogous to the case of the Wiener-Fliess composition when $c$ is restricted to being proper as stated in Theorem~\ref{thm:wiener_fliess_global}. The following theorem asserts that Wiener-Fliess composition preserves global convergence when the commutative series $d$ is restricted to being a polynomial.	
\begth \label{thm:WF_global_p}Let $d \in \allcommutingpolyX{}$ with $\abs{\tilde{X}} = m$ and $c \in \serSinf{m}$, then $d \mixcomp c \in \serSinf{}$.
\endth

\begpr Recall 
\begin{align*}
c \in \serSinf{m} &\Longleftrightarrow c \in \serSinfR{m}{R} \; \forall R > 0.
\end{align*}
Fix $\epsilon \in ]0,1[$ and $\forall R^{\prime} >0$ define $R = \left(\frac{1}{\epsilon}\right) R^{\prime}$. Since $d \in \allcommutingpolyX{}$, applying Theorem~\ref{thm:WF_global_poly} gives $d \mixcomp c \in \serSinfR{}{R^{\prime}}$. Therefore,
\begin{align*}
&d \mixcomp c \in \serSinfR{}{R^{\prime}} \; \forall R^{\prime} > 0 \Longleftrightarrow d\mixcomp c \in \serSinf{} .
\end{align*}
\endpr 

\begin{table}[tb] \label{tab:mixcomp_conv}
	\centering
	\caption{Summary of convergence results for the Wiener-Fliess composition product.}
	{\begin{tabular}{||c|c|c|c||}
			\hline 
			$d$ & $c$ & $d\mixcomp c$ & Theorem \\ \hline 
			$\allcommutingseriesLCX{}$ & $\allseriesLC{m}$ & $\allseriesLC{}$ & Theorem~\ref{thm:loc_wiener}\\ \hline
			$\allcommutingseriesGCX{}$ & $\serSinfR{m}{R}\cap\allproperseries{m}$ & $\serSinfR{}{R^{\prime}}$ where 
			$R^{\prime} = \epsilon R \; \forall \epsilon \in ]0,1[$. & Theorem~\ref{thm:wiener_fliess_global} \\ \hline
			$\allcommutingseriesGCX{}$ & $\serSinf{m}\cap\allproperseries{m}$ & $\serSinf{}$ & Theorem~\ref{thm:WF_glo_conv_ser}  \\ \hline
			$\allcommutingpolyX{}$ & $\allseriesGC{m}$ & $\allseriesGC{}$ & Theorem~\ref{thm:WFcomp_GC}\\ \hline
			$\allcommutingpolyX{}$ & $\serSinfR{m}{R}$ & $\serSinfR{}{R^{\prime}}$ where 
			$R^{\prime} = \epsilon R \; \forall \epsilon \in ]0,1[$. & Theorem~\ref{thm:WF_global_poly} \\ \hline
			$\allcommutingpolyX{}$ & $\serSinf{m}$ & $\serSinf{}$ & Theorem~\ref{thm:WF_global_p}  \\ \hline
	\end{tabular}}
\end{table}

Using Theorem~\ref{thm:finite_radius}, Theorem~\ref{thm:infinite_radius} and Table~$1$, the following conclusions can be drawn. Under the assumptions stated in Theorem~\ref{thm:Wiener-Fliess-product}, the cascade connection of a locally convergent Fliess operator $F_c$ with a locally convergent real analytic function $f_d$ is represented by a locally convergent Fliess operator $F_{d\mixcomp c}$. Similarly, the cascade connection of a globally convergent Fliess operator $F_c$ with a globally convergent real analytic function $f_d$ (as characterized in Theorem~\ref{thm:formal_static_GC}) has a globally convergent Fliess operator representation given by $F_{d\mixcomp c}$. 

This section and Section~\ref{sec:loc_conv} have shown that the Wiener-Fliess composition product preserves both local and global convergence. These results are applied in the next section to prove the local convergence of the Wiener-Fliess feedback product.

\section{Local Convergence of Wiener-Fliess Feedback Product}\label{sec:static_feedback}
\paragraph*{}
The objective of this section is to prove that additive static feedback preserves local convergence. It translates as, a locally convergent Fliess operator $F_c$ in additive static feedback with a locally convergent analytic map $f_d$ is represented by a locally convergent Fliess operator $F_{c\hat{@}d}$. The following is necessary to prove that the  Wiener-Fliess feedback product preserves local convergence.

\begth \label{thm:loc_mod}\cite{GS_thesis} Let $c \in \allseriesLC{}$ such that $\abs{(c,\eta)} \leq K_cM_c^{\abs{\eta}}\abs{\eta}! \;\forall \eta \in X^{\ast}$. Assume $d \in \allseriesLCXtil{m}$ such that $\abs{(d_i,\zeta)} \leq K_{d_i}M_{d_i}^{\abs{\zeta}}\abs{\zeta}!$, where $\zeta \in \tilde{X}^{\ast}$, and $d_i$ is the $i$-th component of the series $d$. Then $c\modcomp d_{\delta} \in \allseriesLCXtil{}$ such that 
\begin{align*}
\abs{(c \modcomp d_{\delta}, \eta)} &\leq \begin{cases}
\left(\dfrac{K_cM_c(1+mK_d)}{(1+mK_d)M_c + M_d}\right) \left[(1+mK_d)M_c + M_d \right]^{\abs{\eta}} \abs{\eta}! &\quad \text{if } \eta \neq \emptyset \\
K_c &\quad \text{if } \eta = \emptyset,
\end{cases}
\end{align*}
where $K_d = \max\limits_{i = 1,\ldots,m}K_{d_i}$ and $M_d = \max\limits_{i = 1,\ldots,m}M_{d_i}$.
\endth

Theorem~\ref{thm:loc_mod} asserts that the mixed composition product preserves local convergence. In addition, the following result is essential to produce the desired result. It states that the antipode of the Hopf algebra corresponding to dynamic output feedback group preserves local convergence.    
\begth\cite{Gray-etal_SCL14}\label{thm:antipode_conv} For any $c \in \allseriesLC{m}$ with growth constants $K_c,M_c > 0$ it follows that
\begin{align*}
\abs{(c^{\circ -1}, \eta)} \leq K(\mathscr{A}(K_c)M_c)^{\abs{\eta}} \abs{\eta}! \quad \forall \eta \in \allwords
\end{align*}
for some $K > 0$ and 
\begin{align*}
\mathscr{A}(K_c)  = \frac{1}{1-mK_c \ln(1 + \frac{1}{mK_c})}.
\end{align*}
\endth 

Theorem~\ref{thm:antipode_conv} is essential as the additive static feedback product involves the antipode operation from the dynamic output feedback group. The following states that the local convergence is preserved by the Wiener-Fliess feedback product.
\begth\label{thm:additive_static_feedb} Given a series $c \in \allseriesLC{\ell}$ and $d \in \allcommutingseriesLCX{m}$ with $\abs{X} = m+1$ and $\abs{\tilde{X}} = \ell$, if either of the following conditions hold:
\begin{enumerate}
	\item The series $c$ is proper,
	\item The commutative series $d$ is a polynomial,
\end{enumerate}
then $c\hat{@}d \in \allseriesLC{\ell}$. Specifically, if $c$ is proper, then $c\hat{@}d \in \allseriespLC{\ell}$.  
\endth

\begpr Consider the case of $c$ being a proper series. Clearly, $d \in \allcommutingseriesLCX{m}$ if and only if $-d \in \allcommutingseriesLCX{m}$. Since $c \in \allseriespLC{\ell}$, then by Theorem \ref{thm:loc_wiener}
\begin{align*}
&(-d \mixcomp c) \in \allseriesLC{m}.
\end{align*}
Hence, applying Theorem \ref{thm:antipode_conv} yields
\begin{align*}
&(-d \mixcomp c)^{\circ -1} \in \allseriesLC{m} \Leftrightarrow \delta + (-d \mixcomp c)^{\circ -1} = (-d \mixcomp c)^{\circ -1}_{\delta} \in \delta+\allseriesLC{m}.
\end{align*}
Therefore, by Theorem \ref{thm:loc_mod}
\begin{align*}
&c\hat{@}d = c \modcomp (-d \mixcomp c)_{\delta}^{\circ -1} \in \allseriesLC{\ell}.
\end{align*}
Since $c$ is proper, by definition of the mixed composition product,
\begdi
c\hat{@}d = c \modcomp (-d \mixcomp c)_{\delta}^{\circ -1} \in \allseriespLC{\ell}.
\enddi
Now consider the case of $d \in \allcommutingpolyX{m}$. Since $c \in \allseriesLC{\ell}$, by Theorem~\ref{thm:WF_global_poly},  $(-d \mixcomp c) \in \allseriesLC{m}$. The rest of the proof is exactly analogous to the previous case.
\endpr

The following example demonstrates that Wiener-Fliess feedback product does not preserve global convergence.
\begex\cite{Thitsa-Gray_SIAM12} Let $X = \{x_0,x_1\}$. Define $c \in \allseries{}$ as $c = x_{1}^{\ast} = \sum_{k=0}^{\infty} x_1^k$. Observe that $c \in \allseriesGC{}$. The Fliess operator $F_c$ describes the input-output behavior of the state space model
\begin{align*}
\dot{z} &= zu, \quad z(0) = 1,\\
y &= z,
\end{align*}
where $z(t), u(t) \in \re$. Define $d \in \re[[w]]$ as the monomial $d = w$. Note that the monomial $d \in \re_{GC}[[w]]$. The Fliess operator $F_{c\hat{@}d}$ describes the closed-loop system of $F_c$ under unity feedback. The zero-input dynamics of the closed-loop system are given by the solution of the following differential equation
\begin{align*}
\dot{z} = z^2, \quad z(0) = 1.
\end{align*}
Specifically, $z(t) = \left(\dfrac{1}{1-t}\right) = 1+t+t^2+\cdots$ for $t < 1$. Recall that $E_{x_0^n}[u] = \dfrac{t^n}{n!}$. The zero-input response therefore corresponds to $F_{\left(c\hat{@}d\right)_N}$, where $\left(c\hat{@}d\right)_N$ is the natural part of the Wiener-Fliess feedback product given by    
\begin{align*}
\left(c \hat{@} d\right)_N = \sum_{k = 0}^{\infty}k! x_{0}^k.
\end{align*}
Observe that the subseries $\left(c\hat{@}d\right)_N$ is only locally convergent. Hence, the Wiener-Fliess feedback product of $c \in \allseriesGC{}$ and $d \in \re_{GC}[[w]]$, $c\hat{@}d$, is only locally convergent. 
\endex

Finally, under the assumptions stated in Theorem~\ref{thm:additive_static_feedb} and applying Theorem~\ref{thm:finite_radius} the following statement can be asserted. The additive static feedback connection of a locally convergent Fliess operator $F_c$ with a locally convergent real analytic function $f_d$ is represented by a locally convergent Fliess operator $F_{c\hat{@}d}$.

\section{Conclusions}
It was shown that the generating series of a closed-loop system consisting of a Chen-Fliess series and a formal static output feedback in an additive configuration always has a Chen-Fliess series representation. To prove this, the generating series of Chen-Fliess series composed with a formal static map was first extensively characterized, including its effect on the relative degree of plant. The computation of the closed-loop generating series was facilitated by the interplay of two Hopf algebras: the Hopf algebra corresponding to the dynamic feedback group and the Hopf algebra corresponding to the shuffle group. The additive static output feedback product was interpreted as a group action on the plant and was also shown to preserve the relative degree of the plant. It was proven to preserve local convergence. A counterexample was provided to show that additive static output feedback does not preserve global convergence in general.

\section*{Acknowledgments}
This work was supported by the National Science Foundation under grant CMMI-1839378. The authors want to acknowledge Dr.~Alexander Schmeding of Nord Universitet for providing his insights and discussions on the topic of Fr\'echet spaces.

\end{document}